\documentclass[twocolumn,floatfix,preprintnumbers,nofootinbib,superscriptaddress]{revtex4-2}

\usepackage{ulem}
\usepackage{mathrsfs}
\usepackage{bm}
\usepackage{times}
\usepackage{amssymb,amsbsy,amsmath,amsfonts}
\usepackage{graphicx}
\usepackage{float}
\usepackage{color}
\usepackage{morefloats}
\usepackage{rotating}
\usepackage{srcltx}
\usepackage{slashed}
\usepackage{subfigure}
\usepackage{multirow}
\usepackage{verbatim}
\usepackage{hyperref}
\usepackage{tabularx}
\usepackage{braket}
\usepackage{diagbox}
\usepackage{appendix}
\usepackage{subfigure}
\usepackage{makecell}
\usepackage{booktabs}
\usepackage{placeins}

\usepackage[outdir=./]{epstopdf}

\begin{document}
\title{Dynamically generated $h_1$ state by the $K^*\bar{K}^*$ interaction and its $K_1(1270)\bar{K}$ and $b_1(1235)\pi$ decays}
\date{\today}

\author{Qing-Hua Shen}~\email{shenqinghua@impcas.ac.cn}
\affiliation{State Key Laboratory of Heavy Ion Science and Technology, Institute of Modern Physics, Chinese Academy of Sciences, Lanzhou 730000, China} 
 \affiliation{School of Physical Science and Technology, Lanzhou University, Lanzhou 730000, China}
\affiliation{School of Nuclear Sciences and Technology, University of Chinese Academy of Sciences, Beijing 101408, China}

 \author{Li-Sheng Geng}~\email{lisheng.geng@buaa.edu.cn}
\affiliation{School of Physics, Beihang University, Beijing 102206, China} \affiliation{Sino-French Carbon Neutrality Research Center, Ecole Centrale de Pekin/School of General Engineering, Beihang University, Beijing 100191, China} \affiliation{Peng Huanwu Collaborative Center for Research and Education, International Institute for Interdisciplinary and Frontiers, Beihang University, Beijing 100191, China} \affiliation{Beijing Key Laboratory of Advanced Nuclear Materials and Physics, Beihang University, Beijing 102206, China}
\affiliation{Southern Center for Nuclear-Science Theory (SCNT), Institute of Modern Physics, Chinese Academy of Sciences, Huizhou 516000, China}

\author{Xiang Liu}~\email{xiangliu@lzu.edu.cn}
\affiliation{School of Physical Science and Technology, Lanzhou University, Lanzhou 730000, China}
\affiliation{Lanzhou Center for Theoretical Physics, Key Laboratory of Theoretical Physics of Gansu Province, Key Laboratory of Quantum Theory and Applications of MoE, Gansu Provincial Research Center for Basic Disciplines of Quantum Physics, Lanzhou University, Lanzhou 730000, China} 
\affiliation{MoE Frontiers Science Center for Rare Isotopes, Lanzhou University, Lanzhou 730000, China}
 \affiliation{Research Center for Hadron and CSR Physics, Lanzhou University and Institute of Modern Physics of CAS, Lanzhou 730000, China}

\author{Ju-Jun Xie}~\email{xiejujun@impcas.ac.cn}
\affiliation{State Key Laboratory of Heavy Ion Science and Technology, Institute of Modern Physics, Chinese Academy of Sciences, Lanzhou 730000, China} 
\affiliation{School of Nuclear Sciences and Technology, University of Chinese Academy of Sciences, Beijing 101408, China}
\affiliation{Southern Center for Nuclear-Science Theory (SCNT), Institute of Modern Physics, Chinese Academy of Sciences, Huizhou 516000, China}
 
 \begin{abstract}
 
We investigate the dynamically generated $h_1$ state with spin-parity $J^P = 1^+$ and a mass around 1790~MeV, arising from the $K^* \bar{K}^*$ interaction within the chiral unitary approach. The partial decay widths into the $K_1(1270)\bar{K}$ and $b_1(1235)\pi$ channels are calculated via a triangular-loop mechanism. In this mechanism, the $h_1$ state couples to $K^* \bar{K}^*$, and final-state interactions between $K^*$ and $\bar{K}^*$ proceed through pseudoscalar-meson exchange, leading to the final states $\bar{K}$ (or $\pi$) and $K_1(1270)$ [or $b_1(1235)$]. We also present the invariant-mass distributions of a vector meson and a pseudoscalar meson originating from the decays of $K_1(1270)$ or $b_1(1235)$, along with the corresponding decay widths. Our results show that these decay widths are all of the order of a few MeV. We hope that future experiments can test the predictions presented here, thereby helping to identify this $h_1$ state.
 
 \end{abstract}
\maketitle

\section{Introduction}

Hadron spectroscopy provides essential insights into the nonperturbative behavior of the strong interaction. Over the past decades, the field has advanced significantly, particularly with the observation of numerous new hadronic states such as the $XYZ$ and $P_c$ particles, which have attracted broad attention. A central goal of modern hadron spectroscopy is the identification of exotic configurations—including multiquark states, glueballs, and hybrids. These newly discovered states offer valuable opportunities to address this challenge.

In this context, the molecular picture has been widely employed to interpret and predict many of these hadrons, where exotic states are understood as bound systems of color-singlet hadrons (for reviews, see e.g., Refs.~\cite{Liu:2013waa,Hosaka:2016pey,Chen:2016qju,Lebed:2016hpi,Guo:2017jvc,Olsen:2017bmm,Brambilla:2019esw,Richard:2016eis,Meng:2022ozq,Chen:2022asf,Liu:2024uxn,Liu:2019zoy}). Theoretically, molecular states are often treated as dynamically generated resonances or bound states emerging from the interactions among their constituent hadrons. A key approach in this direction is the chiral unitary theory (ChUT), which has become a useful tool in the field and has produced many important results (see, e.g., Refs.~\cite{Oller:1997ti,Oset:1997it,Oller:1998hw,Inoue:2001ip,Jido:2003cb,Roca:2005nm,Guo:2006fu,Roca:2006sz,Guo:2006rp,Geng:2006yb,Gamermann:2006nm,Gamermann:2007fi,Molina:2008jw,Geng:2008gx,Geng:2008gx,Sarkar:2010saz,Xiao:2013jla,Liang:2014kra,Zhou:2014ila,Dias:2018qhp,Yu:2019yfr,Wang:2023jeu,Oset:2022xji,Sun:2018zqs,Sakai:2017avl,Dias:2014pva,Molina:2010tx,Molina:2009ct,Oset:2001cn,Sarkar:2004jh}).

In Ref.~\cite{Geng:2008gx}, an $h_1$ state is dynamically generated by the interaction between two vector mesons in the $K^*\bar{K}^*$ channel with strangeness=0, isospin=0, and spin=1. Moreover, the $K^*\bar{K^*}$ is the only one channel in this sector. According to the $K^*\bar{K}^*$ nature of this $h_1$ state, the near threshold enhancement in the $K^*\bar{K^*}$ mass distribution of the process $J/\psi \to \eta K^{*0}\bar{K}^{*0}$ measured by the BES Collaboration~\cite{BES:2009rue} was investigated with the contribution of the $h_1$ state~\cite{Xie:2013ula}, which provides further support for the existence of an $h_1$ state with mass around $1830 \pm 20$ MeV and width about $110 \pm 10$ MeV\footnote{In that reference, the mass and width of the $h_1$ state were evaluated from the module squared of the $K^* \bar{K}^* \to K^*\bar{K}^*$ scattering amplitude.}. Subsequently, Ref.~\cite{Ren:2014ixa} suggests that this $h_1$ state can be searched for via the process $\eta_c / \eta_c(2S) \to \phi K^*\bar{K^*}$ experimentally. 

However, since the mass of this $h_1$ state lies very close to the $K^*\bar{K}^*$ mass threshold, its signal becomes strongly suppressed in the $K^*\bar{K}^*$ invariant-mass spectrum. Moreover, there are no accessible two-pseudoscalar-meson decay channels~\cite{Xie:2013ula}. The pseudoscalar-vector channel is allowed, but these processes involve anomalous couplings, which are generally considered small due to the higher-order nature of the anomalous term in the chiral expansion~\cite{Nagahiro:2008cv}. It is therefore necessary to search for it in alternative decay channels. Along this line, we investigate in this work its decays into $K_1(1270)\bar{K}$ and $b_1(1235)\pi$ channels, where we consider $K_1(1270)$ and $b_1(1235)$ as dynamically generated by the interaction of vector and pseudoscalar mesons~\cite{Roca:2005nm}.

This article is organized as follows. In the next section, we present the $h_1$ state dynamically generated by the interaction of two vector mesons in the sector with strangeness=0, isospin=0, and spin=1, and discuss its dynamic generation. In Sec.~\ref{decay width}, we present the relevant theoretical formalism for the reaction mechanisms of $h_1 \to K_1(1270)\bar{K}$ and $h_1 \to b_1(1235)\pi$, and the numerical results and discussion are given. Finally, we give a summary in Sec.~\ref{summary}.

\section{An $h_1$ state dynamically generated from the $K^*\bar{K^*}$ interaction}~\label{generation}

\begin{figure}[htbp]
    \centering
    \includegraphics[scale=0.7]{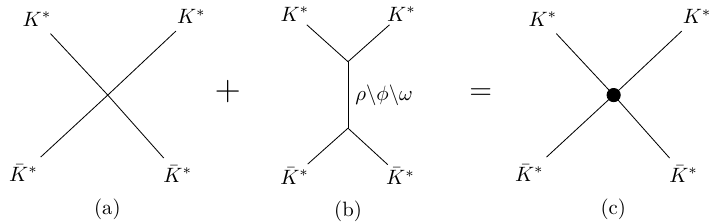}
    \caption{The mechanisms of the $K^*\bar{K}^*$ meson interaction. (a) Four-vector-contact term. (b) $t(u)$-channel vector meson exchange. (c) The effective interaction vertex of $K^*\bar{K}^* \to K^*\bar{K}^*$ at tree level.}
    \label{fig:VV}
\end{figure}

Following Ref.~\cite{Geng:2008gx}, the coupled-channel vector meson-vector meson scattering can be studied with the hidden gauge Lagrangians
    \begin{align}
        \mathcal{L}_{VVVV}&=\frac{1}{2}g^2\langle[V_\mu,V_\nu]V^\mu V^\nu  \rangle, \\
        \mathcal{L}_{VVV}&=ig\langle (V^u \partial_\nu V_\mu-\partial_\nu V_\mu V^\mu)V^\nu\rangle,
    \end{align}
where $g=\frac{M_V}{2f}$, with $M_V = M_\rho$, the mass of the $\rho$ meson, and $f=93$ MeV, the pion decay constant. The $\langle \rangle $ denotes the trace in flavor space. $V_\mu$ stands for the vector nonet, which is given by:
\begin{equation}
    \begin{aligned}
        V_\mu = \begin{pmatrix}
            \frac{1}{\sqrt{2}} (\omega +\rho^0)&\rho^+&K^{*+}\\\rho^-&\frac{1}{\sqrt{2}}(\omega-\rho^0)&K^{*0}\\K^{*-}&\bar{K}^{*0}&\phi
        \end{pmatrix}_\mu
    \end{aligned}.
    \label{eq:Vu}
\end{equation}
For the $K^*\bar{K}^*$ interaction, the domaint mechanisms are shown in Fig.~\ref{fig:VV}, where the four-vector-contact term [Fig.~\ref{fig:VV} (a)] and the $t(u)$-channel vector meson exchange [Fig.~\ref{fig:VV} (b)] are included. Then, one can get the effective interaction vertex, which is shown in Fig.~\ref{fig:VV} (c).

The tree-level $K^*\bar{K}^* \to K^*\bar{K}^*$ transition amplitude $V$ in $s$ wave can be obtained as~\cite{Geng:2008gx}
\begin{equation}
    \begin{aligned}
        V=\frac{g^2[M_\rho^2M_\phi^2+(2M_\rho^2+3M_\phi^2)M_\omega^2](4M_{K^*}^2-3s)}{4M_\rho^2M_\phi^2M_\omega^2},
    \end{aligned}
    \label{eq:V}
\end{equation}
where $s$ is the invariant mass squared of the $K^* \bar{K}^*$ system, and $M_{\rho}$, $M_{\omega}$, $M_{\phi}$, and $M_{K^*}$ stand for the mass of $\rho$, $\omega$, $\phi$, and $K^*$ meson, respectively.
 
Then, we can obtain the unitarized $T$-matrix of the $K^* \bar{K}^* \to K^* \bar{K}^*$ scattering in the $s$ wave as follows,
\begin{equation}
    \begin{aligned}
        T=(1-V\widetilde{G})^{-1}V,
    \end{aligned}
    \label{eq:T}
\end{equation}
where $\widetilde{G}$ is the convoluted loop function to take into account the width of the $K^*$ meson~\cite{Geng:2008gx,Xie:2013ula},
\begin{align}
    \widetilde{G}(s)&=\int_{m_{-}^2}^{m_{+}^2}dm_1^2dm_2^2 A(m_1^2)A(m_2^2)G(s,m_1^2,m_2^2),\\
    A(m^2)&=\frac{1}{N}Im\frac{1}{m^2-m_{K^*}^2+i\Gamma(m^2)m},\\
    N&=\int_{m_{-}^2}^{m_{+}^2}dm^2Im\frac{1}{m^2-m_{K^*}^2+im\Gamma(m^2)},\\
    \Gamma(m^2)&=\Gamma_{K^*}\frac{p^3(m^2)}{p^3(m_{K^*}^2)}\theta (m-m_{K}-m_\pi),\\
    p(m^2)&=\frac{\lambda(m^2,m_\pi^2,m_K^2)}{2m},
    \label{eq:GC} 
\end{align}
where $m_{+}=M_{K^*}+2\Gamma_{K^*}$ and $m_{-} = M_{K^*}-2\Gamma_{K^*}$ are the upper and lower limits of the integration, respectively. The loop function $G$ is given by
\begin{equation}
    \begin{aligned}
        G(s,m_1^2,m_2^2)=i\int\frac{d^4q}{(2\pi)^4}\frac{1}{q-m_1^2+i\epsilon}\frac{1}{(P-q)-m_2^2+i\epsilon}.
    \end{aligned}
\end{equation} 
It is convenient to deal with it by the dimensional regularization method, 
\begin{eqnarray}
 &&   16\pi^2G(s,m_1^2,m_2^2) = a(\mu)+\log\frac{m_1 m_2}{\mu^2}+\frac{\Delta}{2s}\log\frac{m_2^2}{m_1^2} \nonumber \\
 &&   \qquad+ \frac{\nu}{2s}\left(\log\frac{s-\Delta+\nu}{-s+\Delta+\nu}+\log\frac{s+\Delta+\nu}{-s-\Delta+\nu}\right), 
\end{eqnarray}
with $\Delta=m_2^2-m_1^2$, $\nu=\lambda^{\frac{1}{2}}(s,m_1^2,m_2^2)$, and the K$\ddot{a}$ll\'{e}n function $\lambda(x,y,z)=x^2+y^2+z^2-2xy-2yz-2xz$. In addition, $\mu$ is the scale of dimensional regularization, and changes in the scale are reabsorbed in the subtraction constant $a(\mu)$, so that the results remain scale independent. In this work, we take $\mu = 1000$ MeV, as used in previous works~\cite{Geng:2008gx,Xie:2013ula}.

With this formalism and the former ingredients, one can easily obtain the $K^*\bar{K}^* \to K^*\bar{K}^*$ scattering matrix $T$. Then, one can also look for poles of the scattering amplitude $T$ on the complex plane of $s$. The pole $s_{\rm pole}$ on the second Riemann sheet could be associated with the $h_1$ state. The real part of $\sqrt{s_{\rm pole}}$ is associated with the mass of the state, and the minus imaginary part of $\sqrt{s_{\rm pole}}$ is associated with one-half of its width.

\begin{figure}[htbp]
    \centering
    \includegraphics[scale=0.3]{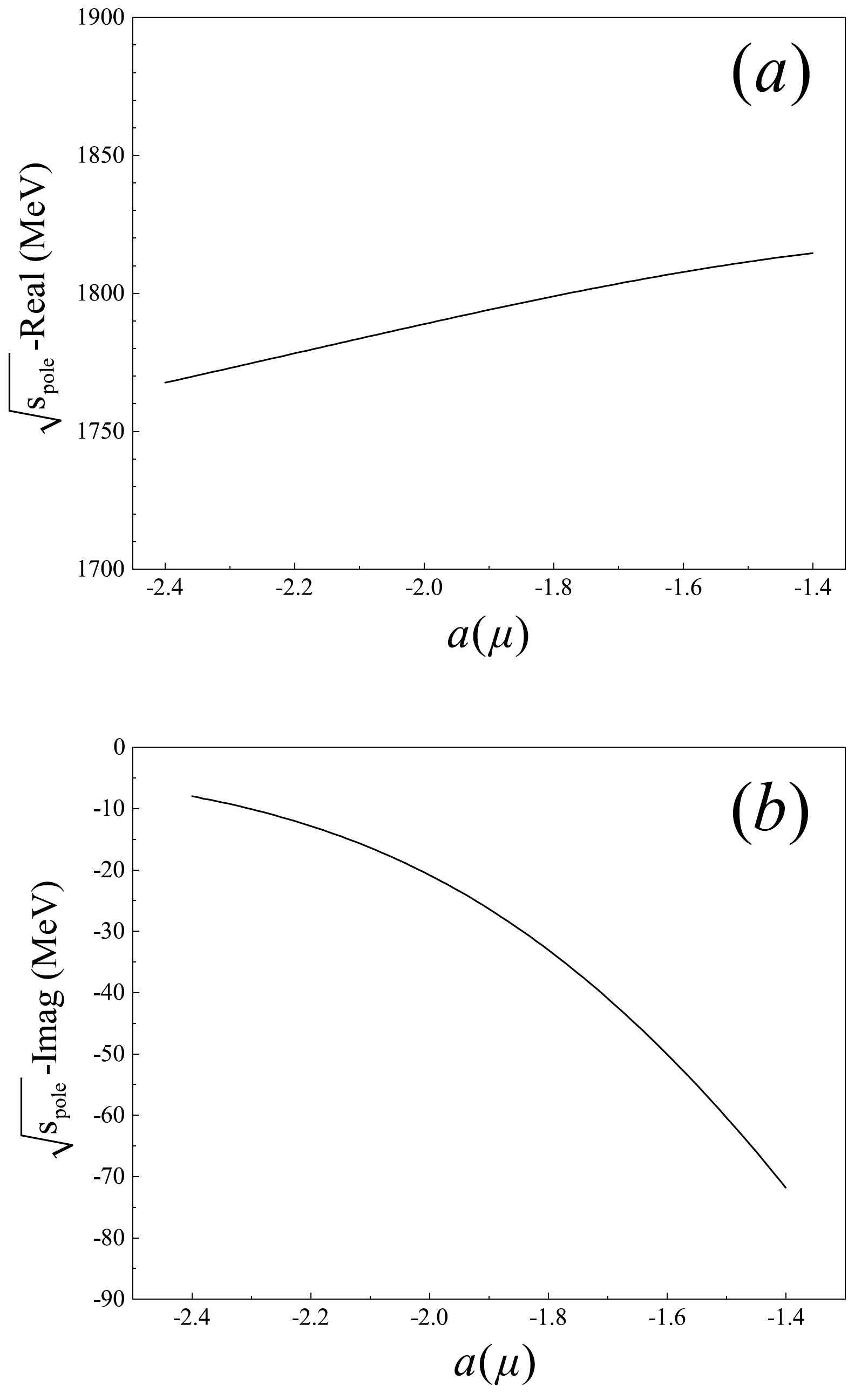}
\caption{The real (a) and imaginary (b) parts of the pole position of the $K^*\bar{K}^* \to K^* \bar{K}^*$ scattering amplitude as a function of the subtraction constant $a(\mu)$.}
    \label{fig:pole(width)}
\end{figure}

In Fig.~\ref{fig:pole(width)}, we show the real and imaginary parts of the pole position of the $h_1$ state as a function of the subtraction constant $a(\mu)$. With the increase of the subtraction constant $a(\mu)$, the mass of $h_1$ becomes larger, and its width becomes broader. Note that, after including the width of $K^*$ and $\bar{K}^*$, in the range of $-2.4$ to $-1.4$ for $a(\mu)$, one can always find a pole for the $h_1$ state with a mass $M_{h_1} = 1790 \pm 25$ MeV, while its width changes much.

It is worth noting that in Ref.~\cite{Xie:2013ula}, the mass and width of the $h_1$ state were evaluated from the squared modulus of the $K^* \bar{K}^* \to K^* \bar{K}^*$ scattering amplitude. The free parameter $a(\mu)$ was determined by fitting to the experimental $K^{*0} \bar{K}^{*0}$ invariant-mass distributions in the $J/\psi \to \eta K^{*0}\bar{K}^{*0}$ decay. Owing to the fact that these experimental data can only be measured above the mass threshold of the $K^{*0} \bar{K}^{*0}$ system, and these data carry large uncertainties, small absolute values of $a(\mu)$ were obtained in that reference. In the present work, we adopt a natural value (around $-2$) for $a(\mu)$, and the mass and width of the $h_1$ state are extracted from the pole of the scattering amplitude $T_{K^* \bar{K}^* \to K^* \bar{K}^*}$ in the complex plane.

\begin{figure}[htbp]
    \centering
    \includegraphics[scale=0.4]{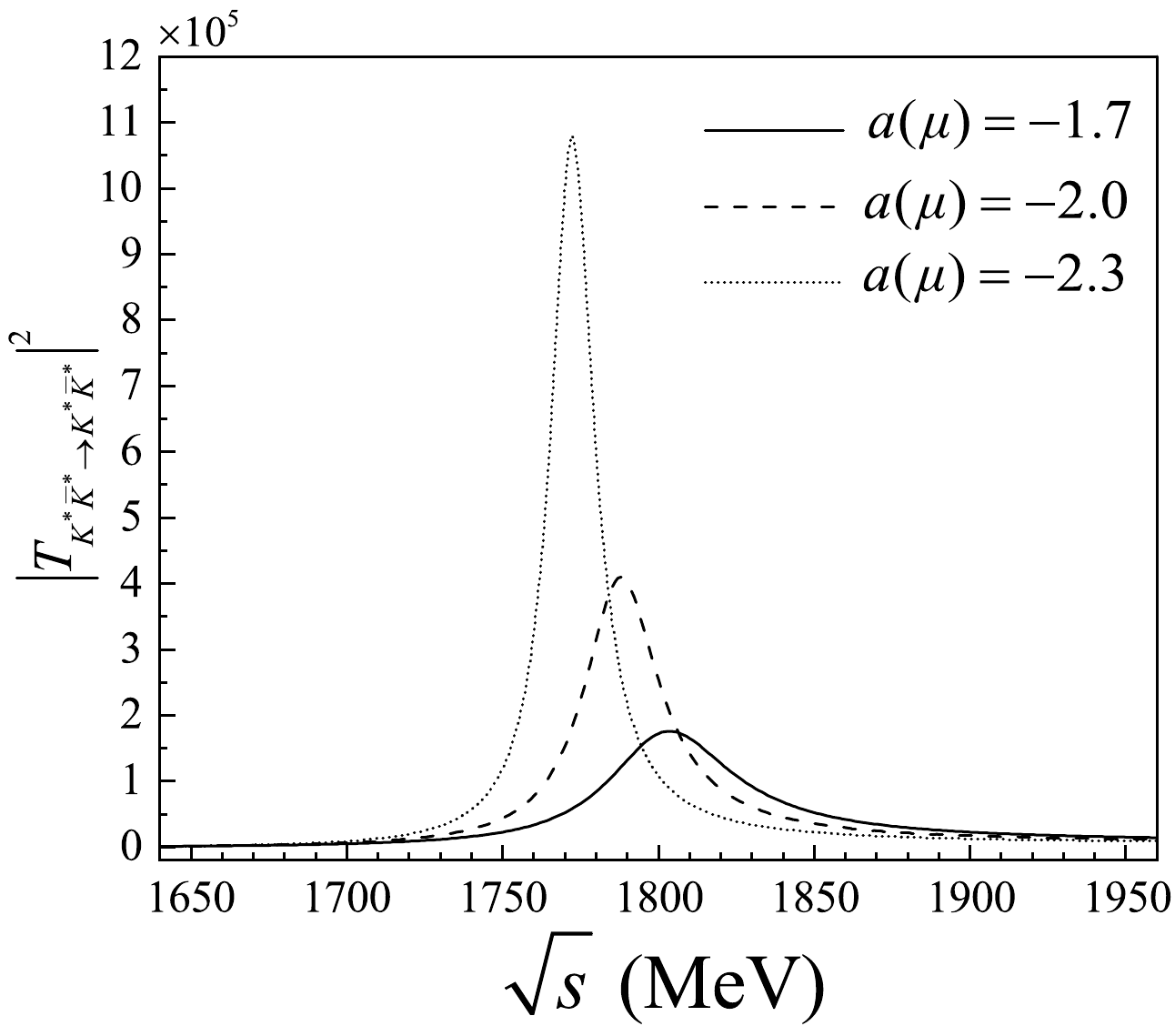}
    \caption{The squared modulus of the $K^*\bar{K}^*$ scattering amplitude in the $I=0$, $S=0$, and $J=1$ sector. The solid, dashed, and densely dotted curves are obtained with the subtraction constant $a(\mu) = -1.7$, $-2.0$, and $-2.3$, respectively.}
    \label{fig:Tsquare}
\end{figure}

In Fig.~\ref{fig:Tsquare}, we present the squared modulus of the $K^*\bar{K}^*$ scattering amplitude in the $I=0$, $S=0$, and $J=1$ sector. The solid, dashed, and densely dotted curves are obtained with the subtraction constant $a(\mu) = -1.7$, $-2.0$, and $-2.3$, respectively. Relative to the real part of scattering amplitude $T_{K^*\bar{K}^* \to K^* \bar{K}^*}$ evaluated in the complex plane, the peak position of $|T_{K^*\bar{K}^* \to K^* \bar{K}^*}|^2$ coincides with that of the real part.

Close to the pole position $s_{\rm pole}$ the scattering amplitude $T$ can be parametrized as the spin projection operator, which projects the vector meson-vector meson pair $K^*\bar{K}^*$ into spin 1 with
\begin{equation}
    \begin{aligned}
        T_{K^* \bar{K}^* \to K^* \bar{K}^*} = \frac{g^2_{h_1K^*\bar{K}^*}}{s-s_{\rm pole}},
    \end{aligned}
\end{equation}
where $g_{h_1K^*\bar{K}^*}$ is the strong coupling constant of the dynamically generated $h_1$ state to the $K^*\bar{K}^*$ channel. Thus, by determining the residues of the scattering amplitude $T_{K^* \bar{K}^* \to K^* \bar{K}^*}$ at the pole position, one can obtain the coupling constant $g_{h_1K^*\bar{K}^*}$, which is complex in general. 

\begin{figure}[htbp]
    \centering
    \includegraphics[scale=0.38]{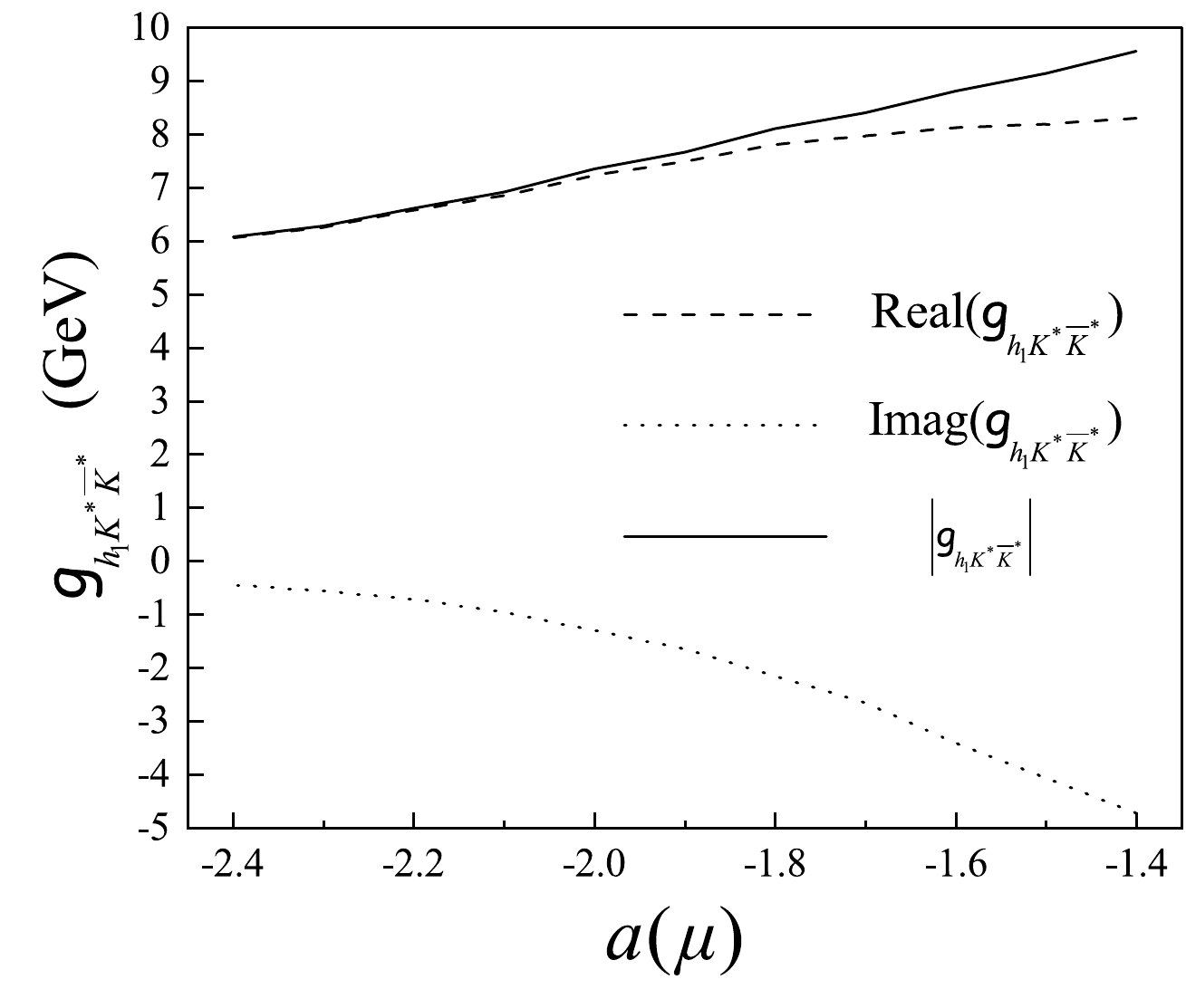}
\caption{Coupling constant of the dynamically generated $h_1$ state to the $K^*\bar{K}^*$ channel as a function of the subtraction constant $a(\mu)$.}
    \label{fig:cp}
\end{figure}

In Fig.~\ref{fig:cp}, we show the theoretical results of $g_{h_1K^*\bar{K}^*}$ (in units of GeV) as a function of $a(\mu)$. It is found that the imaginary part of $g_{h_1K^*\bar{K}^*}$ is small compared with its real part.

\section{The $h_1 \to K_1(1270)\bar{K} \to K^{*+} \pi^0 K^-$ and $h_1 \to b_1(1235)\pi \to \omega(\phi) \pi^+ \pi^-$ decays}~\label{decay width}

\begin{figure}[htbp]
    \centering
    \includegraphics[scale = 0.9]{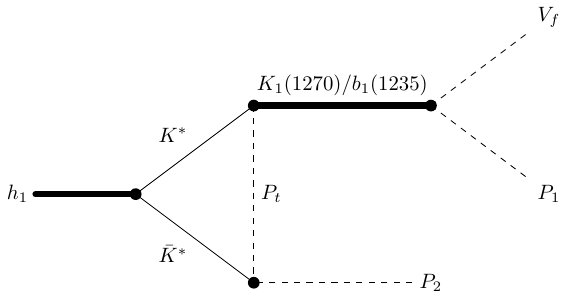}
    \caption{Diagrammatic representation of the triangular-loop mechanism for the $h_1 \to K_1(1270)/b_1(1235) P_2 \to V_f P_1 P_2$ decay. $V_f$, $P_1$ and $P_2$ stand for the vector meson and pseudoscalar mesons in the final state.}     \label{fig:h1AP}
\end{figure}

In this section, we study the $h_1 \to \bar{K} K_1(1270)$ and $\pi b_1(1235)$ decays within a triangular-loop mechanism, where the $h_1$ state couples to the $K^* \bar{K}^*$ channel, and then the final-state interactions between $K^*$ and $\bar{K}^*$ transition to a $\bar{K}$ (or $\pi$) and $K_1(1270)$ [or $b_1(1235)$] through the exchange of pseudoscalar mesons. Here, the axial-vector mesons $K_1(1270)$ and $b_1(1235)$ are viewed as dynamically generated from the vector meson-pseudoscalar meson interactions in coupled channels~\cite{Roca:2005nm}. Thus, the produced $K_1(1270)$ and $b_1(1235)$ resonances then decay into a vector meson plus a pseudoscalar meson. This decay mechanism is illustrated in Fig.~\ref{fig:h1AP}, where $V_f$, $P_1$, and $P_2$ stand for vector mesons and pseudoscalar mesons in the final state. As for the triangle-loop mechanism, several examples exist where similar theoretical analyses have successfully reproduced the experimental data. For example, in Ref.~\cite{Zhang:2017eui}, the partial decay width of $a^+_1(1260) \to \pi^+ \pi^+ \pi^-$ was calculated under the assumption that $a_1(1260)$ is dynamically generated from the coupled-channel $\rho \pi$ and $\bar{K}K^*$ interactions. The resulting $\pi^+\pi^-$ invariant-mass distributions were found to be consistent with experimental data~\cite{ARGUS:1992olh}. In Ref.~~\cite{Xie:2019iwz}, treating $f_1(1285)$ as a dynamically generated state from $\bar{K} K^*$ interaction, the radiative decay rates of $f_1(1285)$ into the $\gamma \rho$, $\gamma \omega$, and $\gamma \phi$ channels were computed by considering the kaon triangle-loop diagrams. The theoretically obtained radiative decay widths were found to be in reasonable agreement with the experimental measurements.

For the $K_1(1270)$ state, it has a two-pole structure in the dynamical generation picture~\cite{Roca:2005nm,Geng:2006yb,Xie:2023cej,Xie:2025xew}. One state corresponds to the lower pole around $(1195-i123)$ MeV, which is referred to as "$K_{1L}(1270)$", while the other one around 1280 MeV is called $K_{1H}(1270)$. The lower one, $K_{1L}(1270)$, mainly couples to the $K^*\pi$ channel. For the $K_{1H}(1270)$, it mainly couples to the $K^* \eta$ channel. Here, we only consider the $K_{1L}(1270)$ state, since the $h_1 \to \bar{K} K_{1H}(1270)$ is suppressed due to the phase space. For the $b_1(1235)$ state, its pole position is $(1247-i28)$ MeV, and it can decay into $\omega \pi$ and $\phi \pi$ channels.

To compute the decay amplitudes of the processes shown in Fig.~\ref{fig:h1AP}, we also need the $AVP$ and $VPP$ interactions, where $A$, $V$, and $P$ stand for axial-vector, vector, and pseudoscalar mesons, respectively. In this work, the effective interaction for the $VPP$ vertex is written as~\cite{Molina:2008jw,Geng:2008gx}
\begin{equation}
    \begin{aligned}
        \mathcal{L}_{VPP}=-ig\langle V_{\mu}[P,\partial^{\mu}P]\rangle. 
    \end{aligned}
    \label{eq:VPP}
\end{equation}

For the $AVP$ vertex, the effective interactions can be written as~\cite{Roca:2006am,Aceti:2015zva,Aceti:2015pma,Xie:2015wja,Molina:2021awn}
\begin{equation}
    \begin{aligned}
        t_{AVP} = g_{AVP} \epsilon_A \cdot \epsilon_V,
    \end{aligned}
    \label{eq:AVP}
\end{equation}
where $\epsilon_A$ and $\epsilon_V$ are the polarization four-vectors of the axial-vector and vector mesons, respectively. Their explicit forms can be found in Refs~\cite{Shen:2024jfr,Gulmez:2016scm}. The coupling constant $g_{AVP}$ can be determined around the pole position of the dynamically generated axial-vector state as discussed before for the case of $g_{h_1 K^* \bar{K^*}}$. We show those involved coupling constants in this work in Table ~\ref{tab:coupling}, which are similar to those obtained in Refs.~\cite{Roca:2005nm,Geng:2006yb}.

\begin{table}[htbp]
     \centering
    \renewcommand\arraystretch{1.5}
    \tabcolsep=16pt
    \caption{The coupling constants (in units of MeV) of the $K_{1L}(1270)$ and $b_1(1235)$ mesons for different channels. }
    \begin{tabular}{c|c}
        \toprule[1pt]
        \toprule[1pt]
        \multicolumn{2}{c}{$K_{1L}(1270) $}\\
        \hline
        $K^*\eta$&$72+i197$\\
        \hline
         $K^*\pi$&$4747-i2874$\\
         \hline
         \multicolumn{2}{c}{$b_1(1235)$}\\
         \hline
         $\frac{1}{\sqrt{2}}(\bar{K}^*K+K^*\bar{K})$&$6172-i75$\\
         \hline
         $\phi \pi$&$2087-i385$\\
         \hline
         $\omega \pi$&$-1869+i300$\\
        \bottomrule[1pt]
        \bottomrule[1pt]
    \end{tabular}
    \label{tab:coupling}
\end{table}

With the formalism and ingredients outlined above, we can write down the decay amplitude for $h_1 \to A P_2 \to V_f P_1 P_2$ depicted in Fig.~\ref{fig:h1AP} as
\begin{eqnarray}
t &=& C gg_{h_1K^*\bar{K}^*}g_{AV_1P_t}g_{AV_fP_1} \int\frac{d^4q}{(2\pi)^4} \\ 
  &&       \times\frac{t_aF^2}{q^2-m_{K^*}^2+im_{K^*}\Gamma_{K^*}} \nonumber \\ \nonumber
  &&       \times \frac{1}{(P-q)^2 - m_{\bar{K}^*}^2 + im_{\bar{K}^*}\Gamma_{\bar{K}^*}} \\ \nonumber
  &&       \times \frac{1}{(q-k_A)^2-m_{P_t}^2+i\epsilon},
    \label{eq:t}
\end{eqnarray}
with
\begin{equation}
    \begin{aligned}
        t_a=& i \sum_{V_A}\mathcal{P}^{(1)*}(K^* \bar{K}^*) \epsilon^*(k_A)\cdot\epsilon(q)\\ &\times \frac{(k_{P_2}+q-k_A)\cdot\epsilon(P-q)  \epsilon(k_A)\cdot \epsilon^*(k_{V_f})}{k^2_A - m^2_A + i m_A\Gamma_A},  \label{eq:ta}
    \end{aligned}
\end{equation}
where $P$, $k_A$, $k_{V_f}$, $k_{p_1}$, and $k_{P_2}$ are the four-momenta of the decay state $h_1$, axial-vector $A$ [$\equiv K_1(1270)$ or $b_1(1235)$], final vector mesons $V_f$, and pseudoscalar mesons $P_1$ and $P_2$, respectively. The $q$ represents the free momentum in the triangular loop. $k_{V_f}$ can be obtained by transforming $k_{V_f}'$ from the rest frame of $V_{f}$ and $P_1$ to the center-of-mass frame of the initial $h_1$ state using Lorentz transformation. In this frame, $k_{V_f}'=(E_{V_f}', |\vec{k}_{V_f}^{ '}|{\rm sin}\alpha {\rm cos}\beta, |\vec{k}_{V_f}^{ '}|{\rm sin}\alpha {\rm sin}\beta, |\vec{k}_{V_f}^{ '}|{\rm cos}\alpha)$, where $\alpha$ and $\beta$ are the polar and azimuth angles, respectively. The energy $E_{V_f}'$ and  momentum $|\vec{k}_{V_f}'|$ is obtained in the center-of-mass frame  of the decaying particle $A$ as
\begin{gather}
E_{V_f}'=\frac{m_A^2 + m_{V_f}^2 - m_{P_1}^2}{2M_A},~~~   |\vec{k}_{V_f}^{ '}|=\sqrt{E_{V_f}'^2 - m_{V_f}^2}.
\end{gather}
The spin projection operator ${\cal P}^{(1)}(K^*\bar{K}^*)$ in Eq.~\eqref{eq:ta} is written as
\begin{equation}
    \begin{aligned}
        \mathcal{P}^{(1)}(K^*\bar{K}^*)=\frac{1}{2}[\epsilon_i(K^*)\epsilon_j(\bar{K}^*)-\epsilon_j(K^*)\epsilon_i(\bar{K}^*)].
    \end{aligned}
\end{equation}

To account for the off-shell effects of the exchanged pseudoscalar meson, a form factor $F$ is introduced~\cite{Molina:2008jw,Geng:2008gx,Wang:2021jub,Wang:2022pin,Titov:2000bn,Titov:2001yw}
\begin{eqnarray}
F &=& \frac{\Lambda_t^2-m_{P_t}^2}{\Lambda_t^2-(E_A - M_{h_1}/2)^2+|\vec{q}-\vec{k}_A|^2}, \\ \label{eq:formfactor}
E_A &=& \frac{M_{h_1}^2 + m^2_A - m^2_{P_2}}{2M_{h_1}},\\
 |\vec{k}_A| &=& \sqrt{E_A^2-m_A^2},
\end{eqnarray}
where $M_{h_1}$ is the mass of the dynamically generated $h_1$ meson, and $E_A$ is the energy of dynamically generated low axial-vector meson $A$. A cutoff parameter $\Lambda_t$ is also included, and we will discuss it later. 

The coefficient $C$ in Eq.~\eqref{eq:t} is an isospin factor, which can be obtained with the following relations~\cite{Dai:2020vfc,Lu:2016nlp,Dai:2018zki}:
\begin{eqnarray}
    \ket{K^*\bar{K}^*}_{(0,0)} &=& -\frac{1}{\sqrt{2}}(\ket{K^{*+}K^{*-}}+\ket{K^{*0}\bar{K}^{*0}}),\\
    \ket{K^*\bar{K}+cc.}_{(1,0)} &=& \frac{1}{2}\left(\ket{\bar{K}^{*0}K^0}-\ket{K^{^*-}K^-}\right.\nonumber\ \\
    && \left.-\ket{K^{*+}K^-}+\ket{K^{*0}\bar{K}^0}\right),\\
     \ket{K^*\bar{K}+cc.}_{(1,-1)} &=& -\frac{1}{\sqrt{2}}(\ket{K^{*-}K^0}+\ket{K^{*0}K^{-}}),\\
      \ket{K^*\bar{K}+cc.}_{(1,1)} &=& \frac{1}{\sqrt{2}}(\ket{K^{*0}K^+}+\ket{K^{*0}K^{*+}}), \\
      \ket{K^*\pi}_{(\frac{1}{2},-\frac{1}{2})} &=& \frac{1}{\sqrt{3}}\ket{K^{*0}\pi^0}-\sqrt{\frac{2}{3}}\ket{K^{*+}\pi^-},\\
      \ket{K^*\pi}_{(\frac{1}{2},\frac{1}{2})} &=& -\sqrt{\frac{2}{3}}\ket{K^{*0}\pi^+}-\frac{1}{\sqrt{3}}\ket{K^{*+}\pi^0}, \\
      \ket{K^{*}\eta}_{(\frac{1}{2},\frac{1}{2})} &=& \ket{K^{*+}\eta},\\
      \ket{K^{*}\eta}_{(\frac{1}{2},-\frac{1}{2})} &=& \ket{K^{*0}\eta}, \\
      \ket{\omega \pi}_{(1,1)} &=& -\ket{\omega \pi^+}, ~~~      \ket{\omega \pi}_{(1,0)} = \ket{\omega \pi^0},\\
      \ket{\omega \pi}_{(1,-1)} &=& \ket{\omega \pi^-}, \\
      \ket{\phi\pi}_{(1,1)} &=& -\ket{\phi \pi^+}, ~~~      \ket{\phi\pi}_{(1,0)} = \ket{\phi \pi^0},\\
      \ket{\phi\pi}_{(1,-1)} &=& \ket{\phi \pi^-},
\end{eqnarray}
where $(\cdot,\cdot)$ presents $(I,I_z)$, where $I$ and $I_z$ are the total isospin and its third component. 

\begin{figure}[htbp]
    \centering
    \includegraphics[scale=0.95]{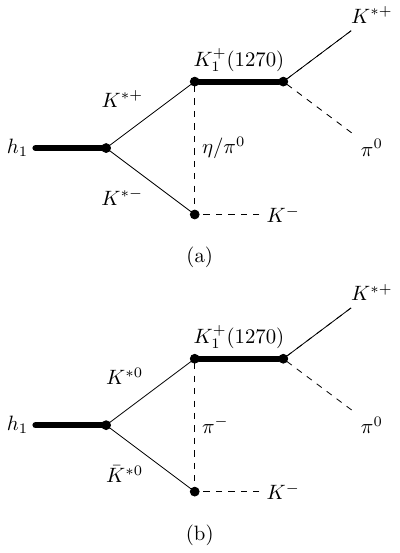}
    %\captionsetup{justification=raggedright, singlelinecheck=false}
    \caption{Diagrammatic representation of the decay mechanisms $h_1 \to K_{1}^+(1270)K^- \to K^{*+}\pi^0 K^-$ through the (a) $K^{*+}-K^{*-}-\eta/\pi^0$ and (b) $K^{*0}-\bar{K}^{*0}-\pi^-$ triangle loops.}
    \label{fig:decay1}
\end{figure}

\begin{figure}[htbp]
    \centering
    \includegraphics[scale=0.95]{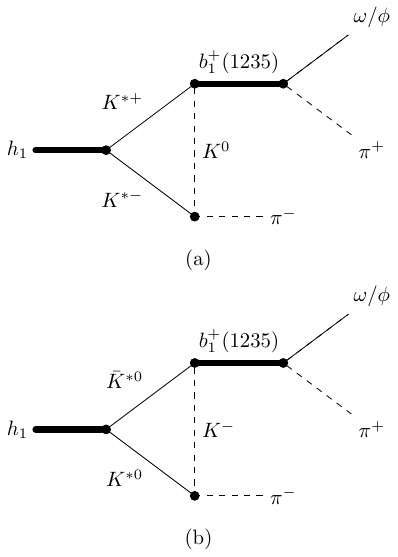}
    %\captionsetup{justification=raggedright, singlelinecheck=false}
    \caption{Diagrammatic representation of the decay mechanisms $h_1 \to b_1^+(1235) \pi^- \to \omega (\phi) \pi^+ \pi^-$ through the (a) $K^{*+}-K^{*-}-K^0$ and $K^{*0}-\bar{K}^{*0}-K^-$ triangle loops.}
    \label{fig:decay2}
\end{figure}

The obtained values of the factor $C$ are shown in Table~\ref{tab:coefficients}. The corresponding decay processes of $h_1 \to K^+_1(1270) K^- \to K^{*+} \pi^0 K^-$ and $h_1 \to b^+_1(1235) \pi^- \to \omega(\phi) \pi^+ \pi^-$ are presented in Figs.~\ref{fig:decay1} and \ref{fig:decay2}, respectively. 

\begin{table}[htbp]
    \centering
    \renewcommand\arraystretch{1.5}
    \tabcolsep=6pt
    \caption{Coefficients $C$ for $h_1 \to A P_2 \to V_f P_1 P_2$ decays.}
    \begin{tabular}{cccccccc}
        \toprule[1pt]
        \toprule[1pt]
        \multicolumn{8}{c}{$h_1 \to K_1^+(1270) K^- \to K^{*+} \pi^0 K^-$}\\
        \hline
         &  & $P_t$ & $A$ & $V_f$ & $P_1$ & $P_2$ & $C$  \\
        \hline
        $K^{*+}$ & $K^{*-}$ & $\eta$ & $K_{1}^+(1270)$ & $K^{*+}$ & $\pi^0$ & $K^-$ & $\frac{1}{2}$\\
        \hline
        $K^{*+}$ & $K^{*-}$ & $\pi^0$ & $K_{1}^+(1270)$ & $K^{*+}$ & $\pi^0$ & $K^-$ & $-\frac{1}{6}$\\
        \hline
        $K^{*0}$ & $\bar{K}^{*0}$ & $\pi^-$ & $K_{1}^+(1270)$ & $K^{*+}$ & $\pi^0$ & $K^-$ & $-\frac{1}{3}$\\
        \hline
         \multicolumn{8}{c}{$h_1 \to b_1^+(1235) \pi^-  \to \omega \pi^+ \pi^-$}\\
         \hline
         $K^{*+}$ & $K^{*-}$ & $K^0$ & $b_1^+(1235)$ & $\omega$ & $\pi^+$ & $\pi^-$ & $\frac{1}{2}$ \\
         \hline
         $\bar{K}^{*0}$ & $K^{*0}$ & $K^-$ & $b_1^+(1235)$ & $\omega$ & $\pi^+$ & $\pi^-$ & $\frac{1}{2}$ \\
         \hline
         \multicolumn{8}{c}{$h_1 \to b_1^+(1235)\pi^- \to \phi \pi^+ \pi^-$}\\
         \hline
          $K^{*+}$ & $K^{*-}$ & $K^0$ & $b_1^+(1235)$ & $\phi$ & $\pi^+$ & $\pi^-$ & $\frac{1}{2}$ \\
         \hline
         $\bar{K}^{*0}$ & $K^{*0}$ & $K^-$ & $b_1^+(1235)$ & $\phi$ & $\pi^+$ & $\pi^-$ & $\frac{1}{2}$ \\
        \bottomrule[1pt]
        \bottomrule[1pt]
    \end{tabular}
    \label{tab:coefficients}
\end{table}

Then, we can obtain the invariant-mass distribution of $V_{f}$ and $P_1$ for the $h_1 \to A P_2 \to V_f P_1 P_2$ decay as,
\begin{equation}
    \begin{aligned}
        \frac{d\Gamma_{h_1 \to A P_2 \to V_f P_1 P_2}}{dM_{V_f P_1}}=\int_{\Omega}\frac{d\Omega}{(2\pi)^4}\frac{|\vec{k}_{A}||\vec{k}_{V_f}'|}{8M_{h_1}^2} \frac{1}{3}\sum_{pol.}|t|^2,
    \end{aligned}
\end{equation}
where we average and sum the spin polarizations of $h_1$ and $V_f$. Here, $\Omega$ is the solid angle in the rest frame of $V_{f}$ and $P_1$, and the differential solid angle is given by $d\Omega = {\rm sin}\alpha d\alpha d\beta$. 
The partial decay width of $h_1 \to A P_2  \to V_f P_1 P_2$ can be also obtained by~\cite{Zeng:2020och}
\begin{equation}
    \begin{aligned}
        \Gamma_{h_1 \to AP_2 \to V_f P_1 P_2} &= \int_{m_{A}-2\Gamma_{A}}^{m_{A}+2\Gamma_{A}}dM_{V_fP_1}  \\
        & \quad\times \frac{d\Gamma_{h_1 \to A P_2 \to V_f P_1 P_2}}{dM_{V_fP_1}},
    \end{aligned}
    \label{eq:width}
\end{equation}
where $m_A$ and $\Gamma_A$ are the mass and width of the dynamically generated axial-vector $A$ mesons, respectively. Here, we take $m_{K_1(1270)} = 1195$, $\Gamma_{K_1(1270)} = 246$, $m_{b_1(1235)} = 1247$, and $\Gamma_{b_1(1235)}= 56$ MeV. 

To calculate the decay amplitude $t$ in Eq.~\eqref{eq:t}, we first integrate over the $q^0$ by the residue theorem, then deal with $|\vec{q}|$ by the cutoff method in the spherical coordinates, where $\Lambda=\Lambda_t=1350\pm100$ MeV is used, which is obtained by matching the experimental data of the two-pseudoscalar-meson decay modes for the dynamically generated scalar and tensor states of two vector mesons~\cite{Shen:2024jfr}.

Additionally, in the calculations, we take
\begin{equation}
    \begin{aligned}
        q &=(q^0,|\vec{q}|{\rm sin}\theta {\rm cos}\phi, |\vec{q}| {\rm sin}\theta {\rm sin}\phi, |\vec{q}|{\rm cos}\theta),\\
        k_A &=(E_A,  |\vec{k_A}|,0,0), \\
        k_{P_2} &=(M_{h_1} - E_A, -|\vec{k}_{A}|,0,0).
    \end{aligned}
\end{equation}

\section{Numerical results and discussions}

\begin{figure}[htbp]
    \centering
    \includegraphics[scale=0.8]{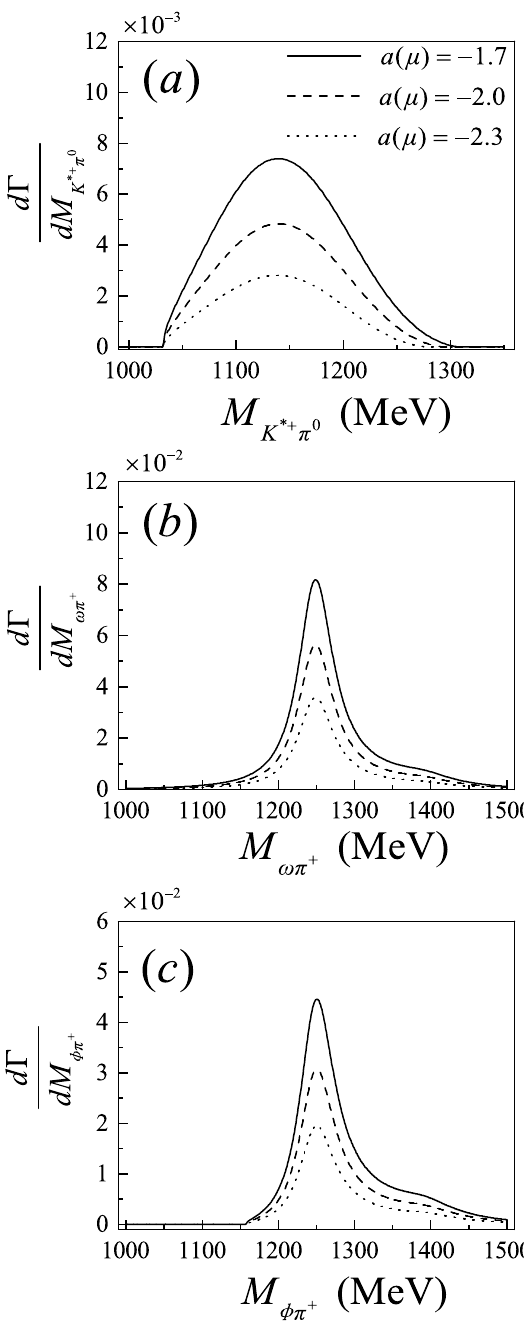}
\caption{The invariant-mass distribution of (a) $K^{*+}\pi^0$ for $h_1 \to K_{1L}^+(1270)K^-,K_{1L}^+(1270) \to K^{*+}\pi^0$, (b) $\omega \pi^+$ for $h_1 \to b_1^+(1235),b_{1}^+(1235) \to \omega\pi^+$, and (c) $\phi \pi^+$ for $h_1 \to b_1^+(1235),b_{1}^+(1235) \to \phi\pi^+$ with $a(\mu) =-1.7$ (solid line), $-2.0$ (dashed line), and $-2.3$ (dotted line).}
    \label{fig:IMD}
\end{figure}

In Fig.~\ref{fig:IMD}, we first show the invariant-mass distribution of $K^{*+}\pi^0$ for the $h_1 \to K_{1}^+(1270)K^- \to K^{*+}\pi^0 K^-$ process [Fig.~\ref{fig:IMD}(a)], $\omega \pi^+$ for $h_1 \to b_1^+(1235) \pi^- \to b_{1}^+(1235) \to \omega\pi^+ \pi^-$ process [Fig.~\ref{fig:IMD}(b)], and $\phi \pi^+$ for $h_1 \to b_1^+(1235) \pi^-  \to \phi \pi^+ \pi^-$ process [Fig.~\ref{fig:IMD}(c)] with $a(\mu)=-1.7$ (solid line), $-2.0$ (dashed line), and $-2.3$ (dotted line), respectively, where the cutoff $\Lambda_t = 1350$ MeV. Note that with different values of $a(\mu)$, the mass of the $h_1$ resonance is also different. For the different cutoffs $\Lambda_t$, the line shapes of these invariant-mass distributions are similar.

In Fig.~\ref{fig:IMD} (a), one can see that the bump structure for the $K_{1L}(1270)$ resonance is lower than the value of the real part of its pole position, which is $1195$ MeV. This is because the mass threshold of $K_{1L}(1270)\bar{K}$ is close to the mass of $h_1$, and the width of $K_{1L}(1270)$ is relatively large. The available phase space for the $h_1 \to K_{lL}(1270) K^- \to K^{*0} \pi^+ K^-$ process is approximately equal to $\Gamma_{K_{1L}(1270)}$. In Figs.~\ref{fig:IMD} (b) and (c), clear peak structures for the $b_1(1235)$ resonance can be observed in the invariant-mass distribution of $\omega \pi^+$ and $\phi \pi^+$, respectively. Additionally, a small bump structure appears around 1400 MeV in Figs.~\ref{fig:IMD} (b) and (c), which originates from a two-body threshold singularity in the triangle diagram as investigated in Ref.~\cite{Bayar:2016ftu}. This singularity occurs when $M_{\omega \pi^+}(M_{\phi\pi^+}) = m_{K^*}+m_{K}$, which is approximately 1400 MeV, and corresponds to the condition $E_A^2 - \omega_{K^*} - \omega_{K}=0$, where $\omega_i = \sqrt{|\vec{p}_i|^2+m_i^2}$ and $\vec{p}_i$ denotes the momentum of particle $i$ with mass $m_i$. Ordinarily, such a singularity can be avoided by deforming the integration contour in the complex plane. However, this is not always possible. As illustrated in Fig.~3 (b) of Ref.~\cite{Bayar:2016ftu}, the integration contour cannot be deformed when $q_{a+}=q_{a-}$, as it becomes pinched between the poles $q_{a+}$ and $q_{a-}$. The definitions of $q_{a+}$ and $q_{a-}$ are given in Ref.~\cite{Bayar:2016ftu}, and further details can be found therein.

\begin{figure}[htbp]
    \centering
    \includegraphics[scale=0.5]{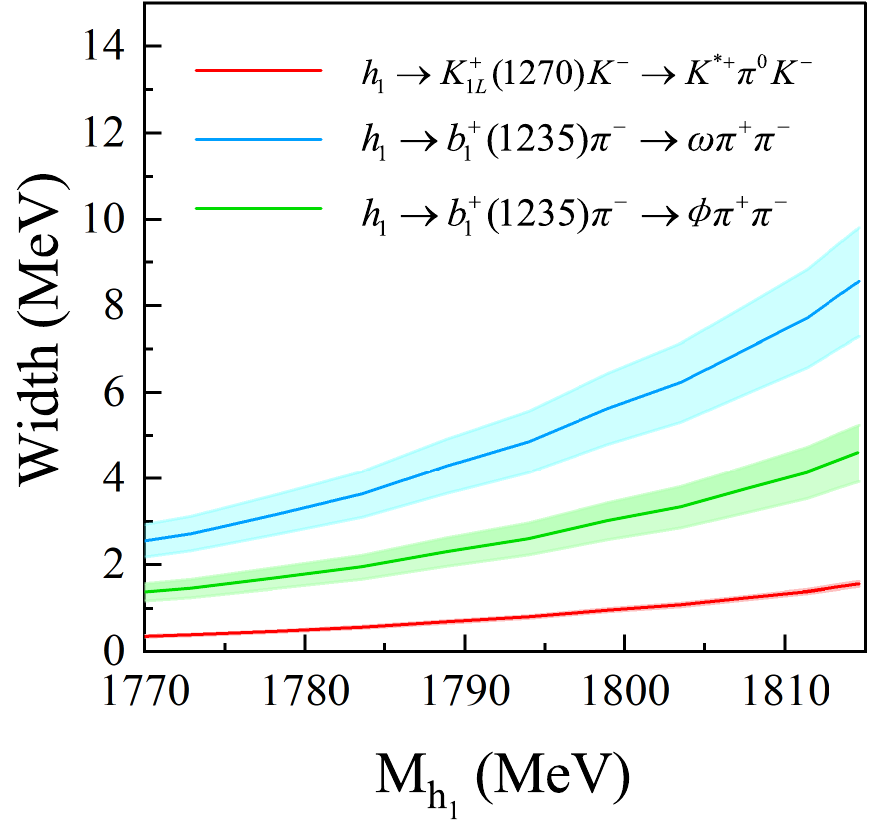}
    \caption{The partial decay width for the processes $h_1 \to K_{1L}^+(1270)K^- \to K^{*+}\pi^0K^-$ (red), $h_1 \to b_1^+(1235) \pi^- \to \omega\pi^+\pi^-$ (blue), and $h_1 \to b_1^+(1235)\pi^0\to \phi\pi^+\pi^-$ (green) with $\Lambda_t = 1350 \pm 100$ MeV.}
    \label{fig:width}
\end{figure}

With Eq.~(\ref{eq:width}), we can calculate the partial decay width of $h_1 \to K_{1}^+(1270)K^- \to K^{*+}\pi^0 K^-$ and $h_1 \to b_1^+(1235) \pi^- \to b_{1}^+(1235) \to \omega (\phi) \pi^+ \pi^-$ processes. The theoretical results are shown in Fig.~\ref{fig:width} with uncertainties from $\Lambda_t = 1350 \pm 100$ MeV. The partial decay width of $h_1 \to K_{1L}^+(1270) K^- \to K^{*+}\pi^0 K^-$ is small compared to the $h_1 \to b_1^+(1235) \pi^- \to  \omega (\phi) \pi^+$ decays, which are on the order of several MeV.

\begin{figure}[htbp]
    \centering
    \includegraphics[scale=0.5]{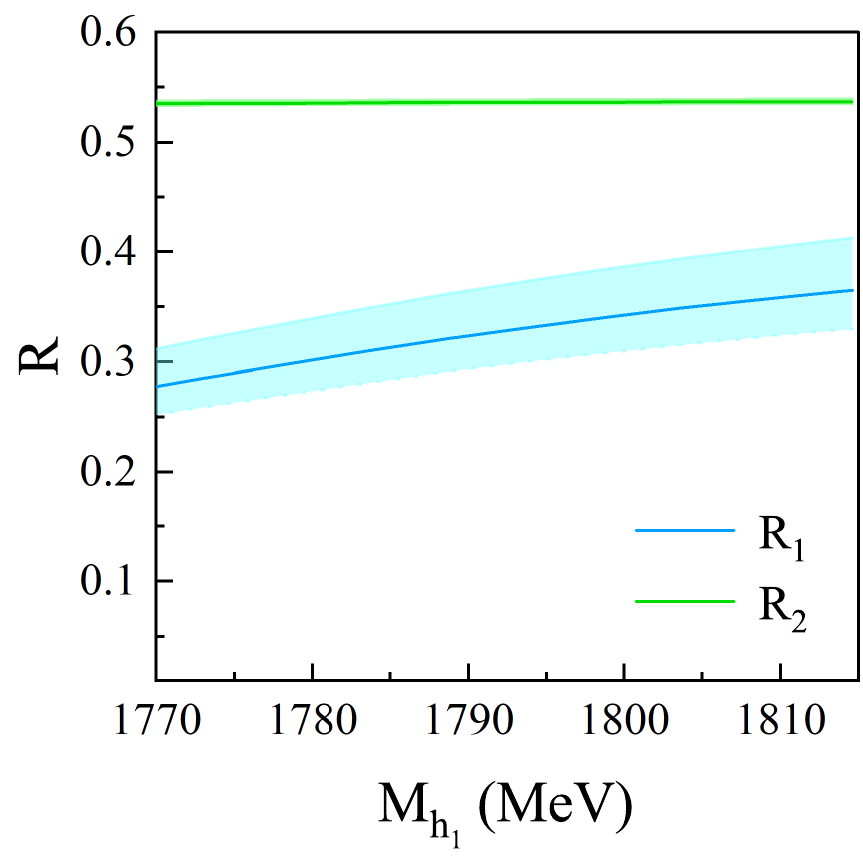}
    \caption{Theoretical ratios of $R_1$ (blue) and $R_2$ (green) for different masses of $h_1$ with $\Lambda_t = 1350 \pm 100$ MeV. }
    \label{fig:R}
\end{figure}

It is known that these partial decay widths are sensitive to the value of the cutoff parameter $\Lambda_t$ and proportional to the product of the coupling constants, which will be canceled in the ratio between different partial decay widths. Therefore, the ratios among these partial decay widths are interesting, and we define 
\begin{equation}
    \begin{aligned}
        R_1&=\frac{\Gamma_{h_1 \to K_{1L}\bar{K},K_{1L} \to K^*\pi}}{\Gamma_{h_1 \to b_1\pi, b_1 \to \omega\pi}},\\
        R_2&=\frac{\Gamma_{h_1 \to b_1\pi, b_1\to \phi\pi}}{\Gamma_{h_1 \to b_1\pi, b_1 \to \omega\pi}},
    \end{aligned}
    \label{eq:R}
\end{equation}
where the $K_{1L}(1270)$ and $b_1(1235)$ are written as $K_{1L}$ and $b_1$. Using the isospin symmetry relations $\Gamma_{h_1 \to K_1^+K^-}/\Gamma_{h_1 \to K_1\bar{K}} = 1/2$ and $\Gamma_{h_1 \to b_1^+\pi^-}/\Gamma_{h_1\to b_1\pi} = 1/3$, the ratio of decay widths for the processes $h_1 \to K_{1L}(1270)\bar{K}$, $K_{1L}(1270) \to K^*\pi$, $h_1 \to b_1(1235)\pi$, $b_1(1235) \to \omega\pi$, and $h_1 \to b_1(1235)\pi$, $b_1(1235) \to \phi\pi$ and the widths of the $h_1 \to K_{1L}^+(1270)K^-$, $K_{1L}^+(1270) \to K^{*+}\pi^0$, $h_1 \to b_1^+(1235) \pi^-$, $b_{1}^+(1235) \to \omega\pi^+$ and $h_1 \to b_1^+(1235) \pi^-$, $b_{1}^+(1235) \to \phi\pi^+$ are 6, 3, and 3, respectively. The ratios $R_1$ (blue) and $R_2$ (green) are shown in Fig.~\ref{fig:R} with $\Lambda_t = 1350 \pm 100$ MeV. Indeed, they are stable with the mass of $M_{h_1}$, and $R_1$ and $R_2$ are approximately 0.3 and 0.53, respectively. We note that only a partial cancellation of the effect from the coupling constant $g_{h_1K^*\bar{K}^*}$ occurs for $R_1$, as clearly seen from the numerical results shown in Fig.~\ref{fig:R}. For $R_2$, contributions from both coupling constants $g_{h_1K^*\bar{K}^*}$ and $g_{b_1 \omega (\phi) \pi}$ are fully canceled.

\section{summary}~\label{summary}

Hadron spectroscopy offers crucial insights into the nonperturbative regime of quantum chromodynamics. In recent years, the observation of numerous unconventional states such as the $XYZ$ and $P_c$ particles has stimulated intense research into exotic hadrons—including multiquark states, glueballs, and hybrids. A prominent theoretical framework for describing such states is the molecular picture, in which hadrons are understood as bound systems of color-singlet hadrons~\cite{Liu:2013waa,Hosaka:2016pey,Chen:2016qju,Lebed:2016hpi,Guo:2017jvc,Olsen:2017bmm,Brambilla:2019esw,Richard:2016eis,Meng:2022ozq,Chen:2022asf,Liu:2024uxn,Liu:2019zoy}. Within this approach, the ChUT has proven to be an effective tool for studying dynamically generated resonances, leading to many successful predictions and interpretations in hadron spectroscopy~\cite{Oller:1997ti,Oset:1997it,Oller:1998hw,Inoue:2001ip,Jido:2003cb,Roca:2005nm,Guo:2006fu,Roca:2006sz,Guo:2006rp,Geng:2006yb,Gamermann:2006nm,Gamermann:2007fi,Molina:2008jw,Geng:2008gx,Geng:2008gx,Sarkar:2010saz,Xiao:2013jla,Liang:2014kra,Zhou:2014ila,Dias:2018qhp,Yu:2019yfr,Wang:2023jeu,Oset:2022xji,Sun:2018zqs,Sakai:2017avl,Dias:2014pva,Molina:2010tx,Molina:2009ct,Oset:2001cn,Sarkar:2004jh}.

In this work, we study the $h_1$ state dynamically generated from the $K^*\bar{K}^*$ interaction within the ChUT, and examine its decays into $K_1(1270)\bar{K}$ and $b_1(1235)\pi$. The partial widths for $h_1 \to K_1^+(1270) K^- \to K^{*+}\pi^0 K^-$ and $h_1 \to b_1^+(1235) \pi^- \to \omega(\phi) \pi^+ \pi^-$ are calculated by including triangle loops involving $K^*$, $\bar{K}^*$, and exchanged pseudoscalar mesons. Here, the axial-vector resonances $K_1(1270)$ and $b_1(1235)$ are themselves treated as dynamically generated states from coupled-channel vector–pseudoscalar interactions.

Within the allowed range of model parameters, the partial decay width for $h_1 \to K_1^+(1270) K^- \to K^{*+}\pi^0 K^-$ is found to be about $0.5\,\text{MeV}$, while those for $h_1 \to b_1^+(1235) \pi^- \to \omega(\phi) \pi^+ \pi^-$ are on the order of several MeV. We also present the invariant-mass distributions for $K^{*0}\pi$ from $K_1(1270)$ decay and for $\omega(\phi)\pi$ from $b_1(1235)$ decay. Furthermore, we explore the dependence of the ratios between different decay modes on the mass of the $h_1$ state—a quantity that could serve as a useful experimental signature. We hope that future experiments will test these predictions, thereby helping to identify the $h_1$ resonance dynamically generated from $K^*\bar{K}^*$ interactions.

\begin{acknowledgments}

This work is partly supported by the National Key R\&D Program of China under Grant No. 2023YFA1606703 and by the National Natural Science Foundation of China under Grants No. 12435007, No. 12361141819, and No. 1252200936. X.L. is also supported by the National Natural Science Foundation of China under Grant No. 12335001 and 12247101, the ``111 Center'' under Grants No. B20063, the Natural Science Foundation of Gansu Province (No. 22JR5RA389, No. 25JRRA799), the fundamental Research Funds for the Central Universities, the project for top-notch innovative talents of Gansu province, and Lanzhou City High-Level Talent Funding.
    
\end{acknowledgments} 

\section{Data Availability}

No data were created or analyzed in this study.

\FloatBarrier

\normalem
\bibliographystyle{apsrev4-1.bst}
\bibliography{reference.bib}

\begin{thebibliography}{66}%
\makeatletter
\providecommand \@ifxundefined [1]{%
 \@ifx{#1\undefined}
}%
\providecommand \@ifnum [1]{%
 \ifnum #1\expandafter \@firstoftwo
 \else \expandafter \@secondoftwo
 \fi
}%
\providecommand \@ifx [1]{%
 \ifx #1\expandafter \@firstoftwo
 \else \expandafter \@secondoftwo
 \fi
}%
\providecommand \natexlab [1]{#1}%
\providecommand \enquote  [1]{``#1''}%
\providecommand \bibnamefont  [1]{#1}%
\providecommand \bibfnamefont [1]{#1}%
\providecommand \citenamefont [1]{#1}%
\providecommand \href@noop [0]{\@secondoftwo}%
\providecommand \href [0]{\begingroup \@sanitize@url \@href}%
\providecommand \@href[1]{\@@startlink{#1}\@@href}%
\providecommand \@@href[1]{\endgroup#1\@@endlink}%
\providecommand \@sanitize@url [0]{\catcode `\\12\catcode `\$12\catcode `\&12\catcode `\#12\catcode `\^12\catcode `\_12\catcode `\%12\relax}%
\providecommand \@@startlink[1]{}%
\providecommand \@@endlink[0]{}%
\providecommand \url  [0]{\begingroup\@sanitize@url \@url }%
\providecommand \@url [1]{\endgroup\@href {#1}{\urlprefix }}%
\providecommand \urlprefix  [0]{URL }%
\providecommand \Eprint [0]{\href }%
\providecommand \doibase [0]{http://dx.doi.org/}%
\providecommand \selectlanguage [0]{\@gobble}%
\providecommand \bibinfo  [0]{\@secondoftwo}%
\providecommand \bibfield  [0]{\@secondoftwo}%
\providecommand \translation [1]{[#1]}%
\providecommand \BibitemOpen [0]{}%
\providecommand \bibitemStop [0]{}%
\providecommand \bibitemNoStop [0]{.\EOS\space}%
\providecommand \EOS [0]{\spacefactor3000\relax}%
\providecommand \BibitemShut  [1]{\csname bibitem#1\endcsname}%
\let\auto@bib@innerbib\@empty
%</preamble>
\bibitem [{\citenamefont {Liu}(2014)}]{Liu:2013waa}%
  \BibitemOpen
  \bibfield  {author} {\bibinfo {author} {\bibfnamefont {X.}~\bibnamefont {Liu}},\ }\href {\doibase 10.1007/s11434-014-0407-2} {\bibfield  {journal} {\bibinfo  {journal} {Chin. Sci. Bull.}\ }\textbf {\bibinfo {volume} {59}},\ \bibinfo {pages} {3815} (\bibinfo {year} {2014})},\ \Eprint {http://arxiv.org/abs/1312.7408} {arXiv:1312.7408 [hep-ph]} \BibitemShut {NoStop}%
\bibitem [{\citenamefont {Hosaka}\ \emph {et~al.}(2016)\citenamefont {Hosaka}, \citenamefont {Iijima}, \citenamefont {Miyabayashi}, \citenamefont {Sakai},\ and\ \citenamefont {Yasui}}]{Hosaka:2016pey}%
  \BibitemOpen
  \bibfield  {author} {\bibinfo {author} {\bibfnamefont {A.}~\bibnamefont {Hosaka}}, \bibinfo {author} {\bibfnamefont {T.}~\bibnamefont {Iijima}}, \bibinfo {author} {\bibfnamefont {K.}~\bibnamefont {Miyabayashi}}, \bibinfo {author} {\bibfnamefont {Y.}~\bibnamefont {Sakai}}, \ and\ \bibinfo {author} {\bibfnamefont {S.}~\bibnamefont {Yasui}},\ }\href {\doibase 10.1093/ptep/ptw045} {\bibfield  {journal} {\bibinfo  {journal} {PTEP}\ }\textbf {\bibinfo {volume} {2016}},\ \bibinfo {pages} {062C01} (\bibinfo {year} {2016})},\ \Eprint {http://arxiv.org/abs/1603.09229} {arXiv:1603.09229 [hep-ph]} \BibitemShut {NoStop}%
\bibitem [{\citenamefont {Chen}\ \emph {et~al.}(2016)\citenamefont {Chen}, \citenamefont {Chen}, \citenamefont {Liu},\ and\ \citenamefont {Zhu}}]{Chen:2016qju}%
  \BibitemOpen
  \bibfield  {author} {\bibinfo {author} {\bibfnamefont {H.-X.}\ \bibnamefont {Chen}}, \bibinfo {author} {\bibfnamefont {W.}~\bibnamefont {Chen}}, \bibinfo {author} {\bibfnamefont {X.}~\bibnamefont {Liu}}, \ and\ \bibinfo {author} {\bibfnamefont {S.-L.}\ \bibnamefont {Zhu}},\ }\href {\doibase 10.1016/j.physrep.2016.05.004} {\bibfield  {journal} {\bibinfo  {journal} {Phys. Rept.}\ }\textbf {\bibinfo {volume} {639}},\ \bibinfo {pages} {1} (\bibinfo {year} {2016})},\ \Eprint {http://arxiv.org/abs/1601.02092} {arXiv:1601.02092 [hep-ph]} \BibitemShut {NoStop}%
\bibitem [{\citenamefont {Lebed}\ \emph {et~al.}(2017)\citenamefont {Lebed}, \citenamefont {Mitchell},\ and\ \citenamefont {Swanson}}]{Lebed:2016hpi}%
  \BibitemOpen
  \bibfield  {author} {\bibinfo {author} {\bibfnamefont {R.~F.}\ \bibnamefont {Lebed}}, \bibinfo {author} {\bibfnamefont {R.~E.}\ \bibnamefont {Mitchell}}, \ and\ \bibinfo {author} {\bibfnamefont {E.~S.}\ \bibnamefont {Swanson}},\ }\href {\doibase 10.1016/j.ppnp.2016.11.003} {\bibfield  {journal} {\bibinfo  {journal} {Prog. Part. Nucl. Phys.}\ }\textbf {\bibinfo {volume} {93}},\ \bibinfo {pages} {143} (\bibinfo {year} {2017})},\ \Eprint {http://arxiv.org/abs/1610.04528} {arXiv:1610.04528 [hep-ph]} \BibitemShut {NoStop}%
\bibitem [{\citenamefont {Guo}\ \emph {et~al.}(2018)\citenamefont {Guo}, \citenamefont {Hanhart}, \citenamefont {Mei{\ss}ner}, \citenamefont {Wang}, \citenamefont {Zhao},\ and\ \citenamefont {Zou}}]{Guo:2017jvc}%
  \BibitemOpen
  \bibfield  {author} {\bibinfo {author} {\bibfnamefont {F.-K.}\ \bibnamefont {Guo}}, \bibinfo {author} {\bibfnamefont {C.}~\bibnamefont {Hanhart}}, \bibinfo {author} {\bibfnamefont {U.-G.}\ \bibnamefont {Mei{\ss}ner}}, \bibinfo {author} {\bibfnamefont {Q.}~\bibnamefont {Wang}}, \bibinfo {author} {\bibfnamefont {Q.}~\bibnamefont {Zhao}}, \ and\ \bibinfo {author} {\bibfnamefont {B.-S.}\ \bibnamefont {Zou}},\ }\href {\doibase 10.1103/RevModPhys.90.015004} {\bibfield  {journal} {\bibinfo  {journal} {Rev. Mod. Phys.}\ }\textbf {\bibinfo {volume} {90}},\ \bibinfo {pages} {015004} (\bibinfo {year} {2018})},\ \bibinfo {note} {[Erratum: Rev.Mod.Phys. 94, 029901 (2022)]},\ \Eprint {http://arxiv.org/abs/1705.00141} {arXiv:1705.00141 [hep-ph]} \BibitemShut {NoStop}%
\bibitem [{\citenamefont {Olsen}\ \emph {et~al.}(2018)\citenamefont {Olsen}, \citenamefont {Skwarnicki},\ and\ \citenamefont {Zieminska}}]{Olsen:2017bmm}%
  \BibitemOpen
  \bibfield  {author} {\bibinfo {author} {\bibfnamefont {S.~L.}\ \bibnamefont {Olsen}}, \bibinfo {author} {\bibfnamefont {T.}~\bibnamefont {Skwarnicki}}, \ and\ \bibinfo {author} {\bibfnamefont {D.}~\bibnamefont {Zieminska}},\ }\href {\doibase 10.1103/RevModPhys.90.015003} {\bibfield  {journal} {\bibinfo  {journal} {Rev. Mod. Phys.}\ }\textbf {\bibinfo {volume} {90}},\ \bibinfo {pages} {015003} (\bibinfo {year} {2018})},\ \Eprint {http://arxiv.org/abs/1708.04012} {arXiv:1708.04012 [hep-ph]} \BibitemShut {NoStop}%
\bibitem [{\citenamefont {Brambilla}\ \emph {et~al.}(2020)\citenamefont {Brambilla}, \citenamefont {Eidelman}, \citenamefont {Hanhart}, \citenamefont {Nefediev}, \citenamefont {Shen}, \citenamefont {Thomas}, \citenamefont {Vairo},\ and\ \citenamefont {Yuan}}]{Brambilla:2019esw}%
  \BibitemOpen
  \bibfield  {author} {\bibinfo {author} {\bibfnamefont {N.}~\bibnamefont {Brambilla}}, \bibinfo {author} {\bibfnamefont {S.}~\bibnamefont {Eidelman}}, \bibinfo {author} {\bibfnamefont {C.}~\bibnamefont {Hanhart}}, \bibinfo {author} {\bibfnamefont {A.}~\bibnamefont {Nefediev}}, \bibinfo {author} {\bibfnamefont {C.-P.}\ \bibnamefont {Shen}}, \bibinfo {author} {\bibfnamefont {C.~E.}\ \bibnamefont {Thomas}}, \bibinfo {author} {\bibfnamefont {A.}~\bibnamefont {Vairo}}, \ and\ \bibinfo {author} {\bibfnamefont {C.-Z.}\ \bibnamefont {Yuan}},\ }\href {\doibase 10.1016/j.physrep.2020.05.001} {\bibfield  {journal} {\bibinfo  {journal} {Phys. Rept.}\ }\textbf {\bibinfo {volume} {873}},\ \bibinfo {pages} {1} (\bibinfo {year} {2020})},\ \Eprint {http://arxiv.org/abs/1907.07583} {arXiv:1907.07583 [hep-ex]} \BibitemShut {NoStop}%
\bibitem [{\citenamefont {Richard}(2016)}]{Richard:2016eis}%
  \BibitemOpen
  \bibfield  {author} {\bibinfo {author} {\bibfnamefont {J.-M.}\ \bibnamefont {Richard}},\ }\href {\doibase 10.1007/s00601-016-1159-0} {\bibfield  {journal} {\bibinfo  {journal} {Few Body Syst.}\ }\textbf {\bibinfo {volume} {57}},\ \bibinfo {pages} {1185} (\bibinfo {year} {2016})},\ \Eprint {http://arxiv.org/abs/1606.08593} {arXiv:1606.08593 [hep-ph]} \BibitemShut {NoStop}%
\bibitem [{\citenamefont {Meng}\ \emph {et~al.}(2023)\citenamefont {Meng}, \citenamefont {Wang}, \citenamefont {Wang},\ and\ \citenamefont {Zhu}}]{Meng:2022ozq}%
  \BibitemOpen
  \bibfield  {author} {\bibinfo {author} {\bibfnamefont {L.}~\bibnamefont {Meng}}, \bibinfo {author} {\bibfnamefont {B.}~\bibnamefont {Wang}}, \bibinfo {author} {\bibfnamefont {G.-J.}\ \bibnamefont {Wang}}, \ and\ \bibinfo {author} {\bibfnamefont {S.-L.}\ \bibnamefont {Zhu}},\ }\href {\doibase 10.1016/j.physrep.2023.04.003} {\bibfield  {journal} {\bibinfo  {journal} {Phys. Rept.}\ }\textbf {\bibinfo {volume} {1019}},\ \bibinfo {pages} {1} (\bibinfo {year} {2023})},\ \Eprint {http://arxiv.org/abs/2204.08716} {arXiv:2204.08716 [hep-ph]} \BibitemShut {NoStop}%
\bibitem [{\citenamefont {Chen}\ \emph {et~al.}(2023)\citenamefont {Chen}, \citenamefont {Chen}, \citenamefont {Liu}, \citenamefont {Liu},\ and\ \citenamefont {Zhu}}]{Chen:2022asf}%
  \BibitemOpen
  \bibfield  {author} {\bibinfo {author} {\bibfnamefont {H.-X.}\ \bibnamefont {Chen}}, \bibinfo {author} {\bibfnamefont {W.}~\bibnamefont {Chen}}, \bibinfo {author} {\bibfnamefont {X.}~\bibnamefont {Liu}}, \bibinfo {author} {\bibfnamefont {Y.-R.}\ \bibnamefont {Liu}}, \ and\ \bibinfo {author} {\bibfnamefont {S.-L.}\ \bibnamefont {Zhu}},\ }\href {\doibase 10.1088/1361-6633/aca3b6} {\bibfield  {journal} {\bibinfo  {journal} {Rept. Prog. Phys.}\ }\textbf {\bibinfo {volume} {86}},\ \bibinfo {pages} {026201} (\bibinfo {year} {2023})},\ \Eprint {http://arxiv.org/abs/2204.02649} {arXiv:2204.02649 [hep-ph]} \BibitemShut {NoStop}%
\bibitem [{\citenamefont {Liu}\ \emph {et~al.}(2025)\citenamefont {Liu}, \citenamefont {Pan}, \citenamefont {Liu}, \citenamefont {Wu}, \citenamefont {Lu},\ and\ \citenamefont {Geng}}]{Liu:2024uxn}%
  \BibitemOpen
  \bibfield  {author} {\bibinfo {author} {\bibfnamefont {M.-Z.}\ \bibnamefont {Liu}}, \bibinfo {author} {\bibfnamefont {Y.-W.}\ \bibnamefont {Pan}}, \bibinfo {author} {\bibfnamefont {Z.-W.}\ \bibnamefont {Liu}}, \bibinfo {author} {\bibfnamefont {T.-W.}\ \bibnamefont {Wu}}, \bibinfo {author} {\bibfnamefont {J.-X.}\ \bibnamefont {Lu}}, \ and\ \bibinfo {author} {\bibfnamefont {L.-S.}\ \bibnamefont {Geng}},\ }\href {\doibase 10.1016/j.physrep.2024.12.001} {\bibfield  {journal} {\bibinfo  {journal} {Phys. Rept.}\ }\textbf {\bibinfo {volume} {1108}},\ \bibinfo {pages} {1} (\bibinfo {year} {2025})},\ \Eprint {http://arxiv.org/abs/2404.06399} {arXiv:2404.06399 [hep-ph]} \BibitemShut {NoStop}%
\bibitem [{\citenamefont {Liu}\ \emph {et~al.}(2019)\citenamefont {Liu}, \citenamefont {Chen}, \citenamefont {Chen}, \citenamefont {Liu},\ and\ \citenamefont {Zhu}}]{Liu:2019zoy}%
  \BibitemOpen
  \bibfield  {author} {\bibinfo {author} {\bibfnamefont {Y.-R.}\ \bibnamefont {Liu}}, \bibinfo {author} {\bibfnamefont {H.-X.}\ \bibnamefont {Chen}}, \bibinfo {author} {\bibfnamefont {W.}~\bibnamefont {Chen}}, \bibinfo {author} {\bibfnamefont {X.}~\bibnamefont {Liu}}, \ and\ \bibinfo {author} {\bibfnamefont {S.-L.}\ \bibnamefont {Zhu}},\ }\href {\doibase 10.1016/j.ppnp.2019.04.003} {\bibfield  {journal} {\bibinfo  {journal} {Prog. Part. Nucl. Phys.}\ }\textbf {\bibinfo {volume} {107}},\ \bibinfo {pages} {237} (\bibinfo {year} {2019})},\ \Eprint {http://arxiv.org/abs/1903.11976} {arXiv:1903.11976 [hep-ph]} \BibitemShut {NoStop}%
\bibitem [{\citenamefont {Oller}\ and\ \citenamefont {Oset}(1997)}]{Oller:1997ti}%
  \BibitemOpen
  \bibfield  {author} {\bibinfo {author} {\bibfnamefont {J.~A.}\ \bibnamefont {Oller}}\ and\ \bibinfo {author} {\bibfnamefont {E.}~\bibnamefont {Oset}},\ }\href {\doibase 10.1016/S0375-9474(97)00160-7} {\bibfield  {journal} {\bibinfo  {journal} {Nucl. Phys. A}\ }\textbf {\bibinfo {volume} {620}},\ \bibinfo {pages} {438} (\bibinfo {year} {1997})},\ \bibinfo {note} {[Erratum: Nucl.Phys.A 652, 407--409 (1999)]},\ \Eprint {http://arxiv.org/abs/hep-ph/9702314} {arXiv:hep-ph/9702314} \BibitemShut {NoStop}%
\bibitem [{\citenamefont {Oset}\ and\ \citenamefont {Ramos}(1998)}]{Oset:1997it}%
  \BibitemOpen
  \bibfield  {author} {\bibinfo {author} {\bibfnamefont {E.}~\bibnamefont {Oset}}\ and\ \bibinfo {author} {\bibfnamefont {A.}~\bibnamefont {Ramos}},\ }\href {\doibase 10.1016/S0375-9474(98)00170-5} {\bibfield  {journal} {\bibinfo  {journal} {Nucl. Phys. A}\ }\textbf {\bibinfo {volume} {635}},\ \bibinfo {pages} {99} (\bibinfo {year} {1998})},\ \Eprint {http://arxiv.org/abs/nucl-th/9711022} {arXiv:nucl-th/9711022} \BibitemShut {NoStop}%
\bibitem [{\citenamefont {Oller}\ \emph {et~al.}(1999)\citenamefont {Oller}, \citenamefont {Oset},\ and\ \citenamefont {Pelaez}}]{Oller:1998hw}%
  \BibitemOpen
  \bibfield  {author} {\bibinfo {author} {\bibfnamefont {J.~A.}\ \bibnamefont {Oller}}, \bibinfo {author} {\bibfnamefont {E.}~\bibnamefont {Oset}}, \ and\ \bibinfo {author} {\bibfnamefont {J.~R.}\ \bibnamefont {Pelaez}},\ }\href {\doibase 10.1103/PhysRevD.59.074001} {\bibfield  {journal} {\bibinfo  {journal} {Phys. Rev. D}\ }\textbf {\bibinfo {volume} {59}},\ \bibinfo {pages} {074001} (\bibinfo {year} {1999})},\ \bibinfo {note} {[Erratum: Phys.Rev.D 60, 099906 (1999), Erratum: Phys.Rev.D 75, 099903 (2007)]},\ \Eprint {http://arxiv.org/abs/hep-ph/9804209} {arXiv:hep-ph/9804209} \BibitemShut {NoStop}%
\bibitem [{\citenamefont {Inoue}\ \emph {et~al.}(2002)\citenamefont {Inoue}, \citenamefont {Oset},\ and\ \citenamefont {Vicente~Vacas}}]{Inoue:2001ip}%
  \BibitemOpen
  \bibfield  {author} {\bibinfo {author} {\bibfnamefont {T.}~\bibnamefont {Inoue}}, \bibinfo {author} {\bibfnamefont {E.}~\bibnamefont {Oset}}, \ and\ \bibinfo {author} {\bibfnamefont {M.~J.}\ \bibnamefont {Vicente~Vacas}},\ }\href {\doibase 10.1103/PhysRevC.65.035204} {\bibfield  {journal} {\bibinfo  {journal} {Phys. Rev. C}\ }\textbf {\bibinfo {volume} {65}},\ \bibinfo {pages} {035204} (\bibinfo {year} {2002})},\ \Eprint {http://arxiv.org/abs/hep-ph/0110333} {arXiv:hep-ph/0110333} \BibitemShut {NoStop}%
\bibitem [{\citenamefont {Jido}\ \emph {et~al.}(2003)\citenamefont {Jido}, \citenamefont {Oller}, \citenamefont {Oset}, \citenamefont {Ramos},\ and\ \citenamefont {Meissner}}]{Jido:2003cb}%
  \BibitemOpen
  \bibfield  {author} {\bibinfo {author} {\bibfnamefont {D.}~\bibnamefont {Jido}}, \bibinfo {author} {\bibfnamefont {J.~A.}\ \bibnamefont {Oller}}, \bibinfo {author} {\bibfnamefont {E.}~\bibnamefont {Oset}}, \bibinfo {author} {\bibfnamefont {A.}~\bibnamefont {Ramos}}, \ and\ \bibinfo {author} {\bibfnamefont {U.~G.}\ \bibnamefont {Meissner}},\ }\href {\doibase 10.1016/S0375-9474(03)01598-7} {\bibfield  {journal} {\bibinfo  {journal} {Nucl. Phys. A}\ }\textbf {\bibinfo {volume} {725}},\ \bibinfo {pages} {181} (\bibinfo {year} {2003})},\ \Eprint {http://arxiv.org/abs/nucl-th/0303062} {arXiv:nucl-th/0303062} \BibitemShut {NoStop}%
\bibitem [{\citenamefont {Roca}\ \emph {et~al.}(2005)\citenamefont {Roca}, \citenamefont {Oset},\ and\ \citenamefont {Singh}}]{Roca:2005nm}%
  \BibitemOpen
  \bibfield  {author} {\bibinfo {author} {\bibfnamefont {L.}~\bibnamefont {Roca}}, \bibinfo {author} {\bibfnamefont {E.}~\bibnamefont {Oset}}, \ and\ \bibinfo {author} {\bibfnamefont {J.}~\bibnamefont {Singh}},\ }\href {\doibase 10.1103/PhysRevD.72.014002} {\bibfield  {journal} {\bibinfo  {journal} {Phys. Rev. D}\ }\textbf {\bibinfo {volume} {72}},\ \bibinfo {pages} {014002} (\bibinfo {year} {2005})},\ \Eprint {http://arxiv.org/abs/hep-ph/0503273} {arXiv:hep-ph/0503273} \BibitemShut {NoStop}%
\bibitem [{\citenamefont {Guo}\ \emph {et~al.}(2006)\citenamefont {Guo}, \citenamefont {Shen}, \citenamefont {Chiang}, \citenamefont {Ping},\ and\ \citenamefont {Zou}}]{Guo:2006fu}%
  \BibitemOpen
  \bibfield  {author} {\bibinfo {author} {\bibfnamefont {F.-K.}\ \bibnamefont {Guo}}, \bibinfo {author} {\bibfnamefont {P.-N.}\ \bibnamefont {Shen}}, \bibinfo {author} {\bibfnamefont {H.-C.}\ \bibnamefont {Chiang}}, \bibinfo {author} {\bibfnamefont {R.-G.}\ \bibnamefont {Ping}}, \ and\ \bibinfo {author} {\bibfnamefont {B.-S.}\ \bibnamefont {Zou}},\ }\href {\doibase 10.1016/j.physletb.2006.08.064} {\bibfield  {journal} {\bibinfo  {journal} {Phys. Lett. B}\ }\textbf {\bibinfo {volume} {641}},\ \bibinfo {pages} {278} (\bibinfo {year} {2006})},\ \Eprint {http://arxiv.org/abs/hep-ph/0603072} {arXiv:hep-ph/0603072} \BibitemShut {NoStop}%
\bibitem [{\citenamefont {Roca}\ \emph {et~al.}(2006)\citenamefont {Roca}, \citenamefont {Sarkar}, \citenamefont {Magas},\ and\ \citenamefont {Oset}}]{Roca:2006sz}%
  \BibitemOpen
  \bibfield  {author} {\bibinfo {author} {\bibfnamefont {L.}~\bibnamefont {Roca}}, \bibinfo {author} {\bibfnamefont {S.}~\bibnamefont {Sarkar}}, \bibinfo {author} {\bibfnamefont {V.~K.}\ \bibnamefont {Magas}}, \ and\ \bibinfo {author} {\bibfnamefont {E.}~\bibnamefont {Oset}},\ }\href {\doibase 10.1103/PhysRevC.73.045208} {\bibfield  {journal} {\bibinfo  {journal} {Phys. Rev. C}\ }\textbf {\bibinfo {volume} {73}},\ \bibinfo {pages} {045208} (\bibinfo {year} {2006})},\ \Eprint {http://arxiv.org/abs/hep-ph/0603222} {arXiv:hep-ph/0603222} \BibitemShut {NoStop}%
\bibitem [{\citenamefont {Guo}\ \emph {et~al.}(2007)\citenamefont {Guo}, \citenamefont {Shen},\ and\ \citenamefont {Chiang}}]{Guo:2006rp}%
  \BibitemOpen
  \bibfield  {author} {\bibinfo {author} {\bibfnamefont {F.-K.}\ \bibnamefont {Guo}}, \bibinfo {author} {\bibfnamefont {P.-N.}\ \bibnamefont {Shen}}, \ and\ \bibinfo {author} {\bibfnamefont {H.-C.}\ \bibnamefont {Chiang}},\ }\href {\doibase 10.1016/j.physletb.2007.01.050} {\bibfield  {journal} {\bibinfo  {journal} {Phys. Lett. B}\ }\textbf {\bibinfo {volume} {647}},\ \bibinfo {pages} {133} (\bibinfo {year} {2007})},\ \Eprint {http://arxiv.org/abs/hep-ph/0610008} {arXiv:hep-ph/0610008} \BibitemShut {NoStop}%
\bibitem [{\citenamefont {Geng}\ \emph {et~al.}(2007)\citenamefont {Geng}, \citenamefont {Oset}, \citenamefont {Roca},\ and\ \citenamefont {Oller}}]{Geng:2006yb}%
  \BibitemOpen
  \bibfield  {author} {\bibinfo {author} {\bibfnamefont {L.~S.}\ \bibnamefont {Geng}}, \bibinfo {author} {\bibfnamefont {E.}~\bibnamefont {Oset}}, \bibinfo {author} {\bibfnamefont {L.}~\bibnamefont {Roca}}, \ and\ \bibinfo {author} {\bibfnamefont {J.~A.}\ \bibnamefont {Oller}},\ }\href {\doibase 10.1103/PhysRevD.75.014017} {\bibfield  {journal} {\bibinfo  {journal} {Phys. Rev. D}\ }\textbf {\bibinfo {volume} {75}},\ \bibinfo {pages} {014017} (\bibinfo {year} {2007})},\ \Eprint {http://arxiv.org/abs/hep-ph/0610217} {arXiv:hep-ph/0610217} \BibitemShut {NoStop}%
\bibitem [{\citenamefont {Gamermann}\ \emph {et~al.}(2007)\citenamefont {Gamermann}, \citenamefont {Oset}, \citenamefont {Strottman},\ and\ \citenamefont {Vicente~Vacas}}]{Gamermann:2006nm}%
  \BibitemOpen
  \bibfield  {author} {\bibinfo {author} {\bibfnamefont {D.}~\bibnamefont {Gamermann}}, \bibinfo {author} {\bibfnamefont {E.}~\bibnamefont {Oset}}, \bibinfo {author} {\bibfnamefont {D.}~\bibnamefont {Strottman}}, \ and\ \bibinfo {author} {\bibfnamefont {M.~J.}\ \bibnamefont {Vicente~Vacas}},\ }\href {\doibase 10.1103/PhysRevD.76.074016} {\bibfield  {journal} {\bibinfo  {journal} {Phys. Rev. D}\ }\textbf {\bibinfo {volume} {76}},\ \bibinfo {pages} {074016} (\bibinfo {year} {2007})},\ \Eprint {http://arxiv.org/abs/hep-ph/0612179} {arXiv:hep-ph/0612179} \BibitemShut {NoStop}%
\bibitem [{\citenamefont {Gamermann}\ and\ \citenamefont {Oset}(2007)}]{Gamermann:2007fi}%
  \BibitemOpen
  \bibfield  {author} {\bibinfo {author} {\bibfnamefont {D.}~\bibnamefont {Gamermann}}\ and\ \bibinfo {author} {\bibfnamefont {E.}~\bibnamefont {Oset}},\ }\href {\doibase 10.1140/epja/i2007-10435-1} {\bibfield  {journal} {\bibinfo  {journal} {Eur. Phys. J. A}\ }\textbf {\bibinfo {volume} {33}},\ \bibinfo {pages} {119} (\bibinfo {year} {2007})},\ \Eprint {http://arxiv.org/abs/0704.2314} {arXiv:0704.2314 [hep-ph]} \BibitemShut {NoStop}%
\bibitem [{\citenamefont {Molina}\ \emph {et~al.}(2008)\citenamefont {Molina}, \citenamefont {Nicmorus},\ and\ \citenamefont {Oset}}]{Molina:2008jw}%
  \BibitemOpen
  \bibfield  {author} {\bibinfo {author} {\bibfnamefont {R.}~\bibnamefont {Molina}}, \bibinfo {author} {\bibfnamefont {D.}~\bibnamefont {Nicmorus}}, \ and\ \bibinfo {author} {\bibfnamefont {E.}~\bibnamefont {Oset}},\ }\href {\doibase 10.1103/PhysRevD.78.114018} {\bibfield  {journal} {\bibinfo  {journal} {Phys. Rev. D}\ }\textbf {\bibinfo {volume} {78}},\ \bibinfo {pages} {114018} (\bibinfo {year} {2008})},\ \Eprint {http://arxiv.org/abs/0809.2233} {arXiv:0809.2233 [hep-ph]} \BibitemShut {NoStop}%
\bibitem [{\citenamefont {Geng}\ and\ \citenamefont {Oset}(2009)}]{Geng:2008gx}%
  \BibitemOpen
  \bibfield  {author} {\bibinfo {author} {\bibfnamefont {L.~S.}\ \bibnamefont {Geng}}\ and\ \bibinfo {author} {\bibfnamefont {E.}~\bibnamefont {Oset}},\ }\href {\doibase 10.1103/PhysRevD.79.074009} {\bibfield  {journal} {\bibinfo  {journal} {Phys. Rev. D}\ }\textbf {\bibinfo {volume} {79}},\ \bibinfo {pages} {074009} (\bibinfo {year} {2009})},\ \Eprint {http://arxiv.org/abs/0812.1199} {arXiv:0812.1199 [hep-ph]} \BibitemShut {NoStop}%
\bibitem [{\citenamefont {Sarkar}\ \emph {et~al.}(2010)\citenamefont {Sarkar}, \citenamefont {Sun}, \citenamefont {Oset},\ and\ \citenamefont {Vicente~Vacas}}]{Sarkar:2010saz}%
  \BibitemOpen
  \bibfield  {author} {\bibinfo {author} {\bibfnamefont {S.}~\bibnamefont {Sarkar}}, \bibinfo {author} {\bibfnamefont {B.-X.}\ \bibnamefont {Sun}}, \bibinfo {author} {\bibfnamefont {E.}~\bibnamefont {Oset}}, \ and\ \bibinfo {author} {\bibfnamefont {M.~J.}\ \bibnamefont {Vicente~Vacas}},\ }\href {\doibase 10.1140/epja/i2010-10956-4} {\bibfield  {journal} {\bibinfo  {journal} {Eur. Phys. J. A}\ }\textbf {\bibinfo {volume} {44}},\ \bibinfo {pages} {431} (\bibinfo {year} {2010})},\ \Eprint {http://arxiv.org/abs/0902.3150} {arXiv:0902.3150 [hep-ph]} \BibitemShut {NoStop}%
\bibitem [{\citenamefont {Xiao}\ and\ \citenamefont {Oset}(2013)}]{Xiao:2013jla}%
  \BibitemOpen
  \bibfield  {author} {\bibinfo {author} {\bibfnamefont {C.~W.}\ \bibnamefont {Xiao}}\ and\ \bibinfo {author} {\bibfnamefont {E.}~\bibnamefont {Oset}},\ }\href {\doibase 10.1140/epja/i2013-13139-y} {\bibfield  {journal} {\bibinfo  {journal} {Eur. Phys. J. A}\ }\textbf {\bibinfo {volume} {49}},\ \bibinfo {pages} {139} (\bibinfo {year} {2013})},\ \Eprint {http://arxiv.org/abs/1305.0786} {arXiv:1305.0786 [hep-ph]} \BibitemShut {NoStop}%
\bibitem [{\citenamefont {Liang}\ \emph {et~al.}(2015)\citenamefont {Liang}, \citenamefont {Uchino}, \citenamefont {Xiao},\ and\ \citenamefont {Oset}}]{Liang:2014kra}%
  \BibitemOpen
  \bibfield  {author} {\bibinfo {author} {\bibfnamefont {W.~H.}\ \bibnamefont {Liang}}, \bibinfo {author} {\bibfnamefont {T.}~\bibnamefont {Uchino}}, \bibinfo {author} {\bibfnamefont {C.~W.}\ \bibnamefont {Xiao}}, \ and\ \bibinfo {author} {\bibfnamefont {E.}~\bibnamefont {Oset}},\ }\href {\doibase 10.1140/epja/i2015-15016-1} {\bibfield  {journal} {\bibinfo  {journal} {Eur. Phys. J. A}\ }\textbf {\bibinfo {volume} {51}},\ \bibinfo {pages} {16} (\bibinfo {year} {2015})},\ \Eprint {http://arxiv.org/abs/1402.5293} {arXiv:1402.5293 [hep-ph]} \BibitemShut {NoStop}%
\bibitem [{\citenamefont {Zhou}\ \emph {et~al.}(2014)\citenamefont {Zhou}, \citenamefont {Ren}, \citenamefont {Chen},\ and\ \citenamefont {Geng}}]{Zhou:2014ila}%
  \BibitemOpen
  \bibfield  {author} {\bibinfo {author} {\bibfnamefont {Y.}~\bibnamefont {Zhou}}, \bibinfo {author} {\bibfnamefont {X.-L.}\ \bibnamefont {Ren}}, \bibinfo {author} {\bibfnamefont {H.-X.}\ \bibnamefont {Chen}}, \ and\ \bibinfo {author} {\bibfnamefont {L.-S.}\ \bibnamefont {Geng}},\ }\href {\doibase 10.1103/PhysRevD.90.014020} {\bibfield  {journal} {\bibinfo  {journal} {Phys. Rev. D}\ }\textbf {\bibinfo {volume} {90}},\ \bibinfo {pages} {014020} (\bibinfo {year} {2014})},\ \Eprint {http://arxiv.org/abs/1404.6847} {arXiv:1404.6847 [nucl-th]} \BibitemShut {NoStop}%
\bibitem [{\citenamefont {Dias}\ \emph {et~al.}(2018)\citenamefont {Dias}, \citenamefont {Debastiani}, \citenamefont {Xie},\ and\ \citenamefont {Oset}}]{Dias:2018qhp}%
  \BibitemOpen
  \bibfield  {author} {\bibinfo {author} {\bibfnamefont {J.~M.}\ \bibnamefont {Dias}}, \bibinfo {author} {\bibfnamefont {V.~R.}\ \bibnamefont {Debastiani}}, \bibinfo {author} {\bibfnamefont {J.~J.}\ \bibnamefont {Xie}}, \ and\ \bibinfo {author} {\bibfnamefont {E.}~\bibnamefont {Oset}},\ }\href {\doibase 10.1103/PhysRevD.98.094017} {\bibfield  {journal} {\bibinfo  {journal} {Phys. Rev. D}\ }\textbf {\bibinfo {volume} {98}},\ \bibinfo {pages} {094017} (\bibinfo {year} {2018})},\ \Eprint {http://arxiv.org/abs/1805.03286} {arXiv:1805.03286 [hep-ph]} \BibitemShut {NoStop}%
\bibitem [{\citenamefont {Yu}\ \emph {et~al.}(2019)\citenamefont {Yu}, \citenamefont {Dias}, \citenamefont {Liang},\ and\ \citenamefont {Oset}}]{Yu:2019yfr}%
  \BibitemOpen
  \bibfield  {author} {\bibinfo {author} {\bibfnamefont {Q.-X.}\ \bibnamefont {Yu}}, \bibinfo {author} {\bibfnamefont {J.~M.}\ \bibnamefont {Dias}}, \bibinfo {author} {\bibfnamefont {W.-H.}\ \bibnamefont {Liang}}, \ and\ \bibinfo {author} {\bibfnamefont {E.}~\bibnamefont {Oset}},\ }\href {\doibase 10.1140/epjc/s10052-019-7543-4} {\bibfield  {journal} {\bibinfo  {journal} {Eur. Phys. J. C}\ }\textbf {\bibinfo {volume} {79}},\ \bibinfo {pages} {1025} (\bibinfo {year} {2019})},\ \Eprint {http://arxiv.org/abs/1909.13449} {arXiv:1909.13449 [hep-ph]} \BibitemShut {NoStop}%
\bibitem [{\citenamefont {Wang}\ and\ \citenamefont {Sun}(2023)}]{Wang:2023jeu}%
  \BibitemOpen
  \bibfield  {author} {\bibinfo {author} {\bibfnamefont {Z.-Y.}\ \bibnamefont {Wang}}\ and\ \bibinfo {author} {\bibfnamefont {Z.-F.}\ \bibnamefont {Sun}},\ }\href {\doibase 10.1140/epjc/s10052-023-12283-3} {\bibfield  {journal} {\bibinfo  {journal} {Eur. Phys. J. C}\ }\textbf {\bibinfo {volume} {83}},\ \bibinfo {pages} {1106} (\bibinfo {year} {2023})},\ \Eprint {http://arxiv.org/abs/2307.00803} {arXiv:2307.00803 [hep-ph]} \BibitemShut {NoStop}%
\bibitem [{\citenamefont {Oset}\ and\ \citenamefont {Roca}(2022)}]{Oset:2022xji}%
  \BibitemOpen
  \bibfield  {author} {\bibinfo {author} {\bibfnamefont {E.}~\bibnamefont {Oset}}\ and\ \bibinfo {author} {\bibfnamefont {L.}~\bibnamefont {Roca}},\ }\href {\doibase 10.1140/epjc/s10052-022-10850-8} {\bibfield  {journal} {\bibinfo  {journal} {Eur. Phys. J. C}\ }\textbf {\bibinfo {volume} {82}},\ \bibinfo {pages} {882} (\bibinfo {year} {2022})},\ \bibinfo {note} {[Erratum: Eur.Phys.J.C 82, 1014 (2022)]},\ \Eprint {http://arxiv.org/abs/2207.08538} {arXiv:2207.08538 [hep-ph]} \BibitemShut {NoStop}%
\bibitem [{\citenamefont {Sun}\ \emph {et~al.}(2018)\citenamefont {Sun}, \citenamefont {Xie},\ and\ \citenamefont {Oset}}]{Sun:2018zqs}%
  \BibitemOpen
  \bibfield  {author} {\bibinfo {author} {\bibfnamefont {Z.-F.}\ \bibnamefont {Sun}}, \bibinfo {author} {\bibfnamefont {J.-J.}\ \bibnamefont {Xie}}, \ and\ \bibinfo {author} {\bibfnamefont {E.}~\bibnamefont {Oset}},\ }\href {\doibase 10.1103/PhysRevD.97.094031} {\bibfield  {journal} {\bibinfo  {journal} {Phys. Rev. D}\ }\textbf {\bibinfo {volume} {97}},\ \bibinfo {pages} {094031} (\bibinfo {year} {2018})},\ \Eprint {http://arxiv.org/abs/1801.04367} {arXiv:1801.04367 [hep-ph]} \BibitemShut {NoStop}%
\bibitem [{\citenamefont {Sakai}\ \emph {et~al.}(2017)\citenamefont {Sakai}, \citenamefont {Roca},\ and\ \citenamefont {Oset}}]{Sakai:2017avl}%
  \BibitemOpen
  \bibfield  {author} {\bibinfo {author} {\bibfnamefont {S.}~\bibnamefont {Sakai}}, \bibinfo {author} {\bibfnamefont {L.}~\bibnamefont {Roca}}, \ and\ \bibinfo {author} {\bibfnamefont {E.}~\bibnamefont {Oset}},\ }\href {\doibase 10.1103/PhysRevD.96.054023} {\bibfield  {journal} {\bibinfo  {journal} {Phys. Rev. D}\ }\textbf {\bibinfo {volume} {96}},\ \bibinfo {pages} {054023} (\bibinfo {year} {2017})},\ \Eprint {http://arxiv.org/abs/1704.02196} {arXiv:1704.02196 [hep-ph]} \BibitemShut {NoStop}%
\bibitem [{\citenamefont {Dias}\ \emph {et~al.}(2015)\citenamefont {Dias}, \citenamefont {Aceti},\ and\ \citenamefont {Oset}}]{Dias:2014pva}%
  \BibitemOpen
  \bibfield  {author} {\bibinfo {author} {\bibfnamefont {J.~M.}\ \bibnamefont {Dias}}, \bibinfo {author} {\bibfnamefont {F.}~\bibnamefont {Aceti}}, \ and\ \bibinfo {author} {\bibfnamefont {E.}~\bibnamefont {Oset}},\ }\href {\doibase 10.1103/PhysRevD.91.076001} {\bibfield  {journal} {\bibinfo  {journal} {Phys. Rev. D}\ }\textbf {\bibinfo {volume} {91}},\ \bibinfo {pages} {076001} (\bibinfo {year} {2015})},\ \Eprint {http://arxiv.org/abs/1410.1785} {arXiv:1410.1785 [hep-ph]} \BibitemShut {NoStop}%
\bibitem [{\citenamefont {Molina}\ \emph {et~al.}(2010)\citenamefont {Molina}, \citenamefont {Branz},\ and\ \citenamefont {Oset}}]{Molina:2010tx}%
  \BibitemOpen
  \bibfield  {author} {\bibinfo {author} {\bibfnamefont {R.}~\bibnamefont {Molina}}, \bibinfo {author} {\bibfnamefont {T.}~\bibnamefont {Branz}}, \ and\ \bibinfo {author} {\bibfnamefont {E.}~\bibnamefont {Oset}},\ }\href {\doibase 10.1103/PhysRevD.82.014010} {\bibfield  {journal} {\bibinfo  {journal} {Phys. Rev. D}\ }\textbf {\bibinfo {volume} {82}},\ \bibinfo {pages} {014010} (\bibinfo {year} {2010})},\ \Eprint {http://arxiv.org/abs/1005.0335} {arXiv:1005.0335 [hep-ph]} \BibitemShut {NoStop}%
\bibitem [{\citenamefont {Molina}\ and\ \citenamefont {Oset}(2009)}]{Molina:2009ct}%
  \BibitemOpen
  \bibfield  {author} {\bibinfo {author} {\bibfnamefont {R.}~\bibnamefont {Molina}}\ and\ \bibinfo {author} {\bibfnamefont {E.}~\bibnamefont {Oset}},\ }\href {\doibase 10.1103/PhysRevD.80.114013} {\bibfield  {journal} {\bibinfo  {journal} {Phys. Rev. D}\ }\textbf {\bibinfo {volume} {80}},\ \bibinfo {pages} {114013} (\bibinfo {year} {2009})},\ \Eprint {http://arxiv.org/abs/0907.3043} {arXiv:0907.3043 [hep-ph]} \BibitemShut {NoStop}%
\bibitem [{\citenamefont {Oset}\ \emph {et~al.}(2002)\citenamefont {Oset}, \citenamefont {Ramos},\ and\ \citenamefont {Bennhold}}]{Oset:2001cn}%
  \BibitemOpen
  \bibfield  {author} {\bibinfo {author} {\bibfnamefont {E.}~\bibnamefont {Oset}}, \bibinfo {author} {\bibfnamefont {A.}~\bibnamefont {Ramos}}, \ and\ \bibinfo {author} {\bibfnamefont {C.}~\bibnamefont {Bennhold}},\ }\href {\doibase 10.1016/S0370-2693(01)01523-4} {\bibfield  {journal} {\bibinfo  {journal} {Phys. Lett. B}\ }\textbf {\bibinfo {volume} {527}},\ \bibinfo {pages} {99} (\bibinfo {year} {2002})},\ \bibinfo {note} {[Erratum: Phys.Lett.B 530, 260--260 (2002)]},\ \Eprint {http://arxiv.org/abs/nucl-th/0109006} {arXiv:nucl-th/0109006} \BibitemShut {NoStop}%
\bibitem [{\citenamefont {Sarkar}\ \emph {et~al.}(2005)\citenamefont {Sarkar}, \citenamefont {Oset},\ and\ \citenamefont {Vicente~Vacas}}]{Sarkar:2004jh}%
  \BibitemOpen
  \bibfield  {author} {\bibinfo {author} {\bibfnamefont {S.}~\bibnamefont {Sarkar}}, \bibinfo {author} {\bibfnamefont {E.}~\bibnamefont {Oset}}, \ and\ \bibinfo {author} {\bibfnamefont {M.~J.}\ \bibnamefont {Vicente~Vacas}},\ }\href {\doibase 10.1016/j.nuclphysa.2005.01.006} {\bibfield  {journal} {\bibinfo  {journal} {Nucl. Phys. A}\ }\textbf {\bibinfo {volume} {750}},\ \bibinfo {pages} {294} (\bibinfo {year} {2005})},\ \bibinfo {note} {[Erratum: Nucl.Phys.A 780, 90--90 (2006)]},\ \Eprint {http://arxiv.org/abs/nucl-th/0407025} {arXiv:nucl-th/0407025} \BibitemShut {NoStop}%
\bibitem [{\citenamefont {Ablikim}\ \emph {et~al.}(2010)\citenamefont {Ablikim} \emph {et~al.}}]{BES:2009rue}%
  \BibitemOpen
  \bibfield  {author} {\bibinfo {author} {\bibfnamefont {M.}~\bibnamefont {Ablikim}} \emph {et~al.} (\bibinfo {collaboration} {BES}),\ }\href {\doibase 10.1016/j.physletb.2010.01.063} {\bibfield  {journal} {\bibinfo  {journal} {Phys. Lett. B}\ }\textbf {\bibinfo {volume} {685}},\ \bibinfo {pages} {27} (\bibinfo {year} {2010})},\ \Eprint {http://arxiv.org/abs/0909.2087} {arXiv:0909.2087 [hep-ex]} \BibitemShut {NoStop}%
\bibitem [{\citenamefont {Xie}\ \emph {et~al.}(2014)\citenamefont {Xie}, \citenamefont {Albaladejo},\ and\ \citenamefont {Oset}}]{Xie:2013ula}%
  \BibitemOpen
  \bibfield  {author} {\bibinfo {author} {\bibfnamefont {J.-J.}\ \bibnamefont {Xie}}, \bibinfo {author} {\bibfnamefont {M.}~\bibnamefont {Albaladejo}}, \ and\ \bibinfo {author} {\bibfnamefont {E.}~\bibnamefont {Oset}},\ }\href {\doibase 10.1016/j.physletb.2013.12.015} {\bibfield  {journal} {\bibinfo  {journal} {Phys. Lett. B}\ }\textbf {\bibinfo {volume} {728}},\ \bibinfo {pages} {319} (\bibinfo {year} {2014})},\ \Eprint {http://arxiv.org/abs/1306.6594} {arXiv:1306.6594 [hep-ph]} \BibitemShut {NoStop}%
\bibitem [{\citenamefont {Ren}\ \emph {et~al.}(2014)\citenamefont {Ren}, \citenamefont {Geng}, \citenamefont {Oset},\ and\ \citenamefont {Meng}}]{Ren:2014ixa}%
  \BibitemOpen
  \bibfield  {author} {\bibinfo {author} {\bibfnamefont {X.-L.}\ \bibnamefont {Ren}}, \bibinfo {author} {\bibfnamefont {L.-S.}\ \bibnamefont {Geng}}, \bibinfo {author} {\bibfnamefont {E.}~\bibnamefont {Oset}}, \ and\ \bibinfo {author} {\bibfnamefont {J.}~\bibnamefont {Meng}},\ }\href {\doibase 10.1140/epja/i2014-14133-7} {\bibfield  {journal} {\bibinfo  {journal} {Eur. Phys. J. A}\ }\textbf {\bibinfo {volume} {50}},\ \bibinfo {pages} {133} (\bibinfo {year} {2014})},\ \Eprint {http://arxiv.org/abs/1405.0153} {arXiv:1405.0153 [nucl-th]} \BibitemShut {NoStop}%
\bibitem [{\citenamefont {Nagahiro}\ \emph {et~al.}(2009)\citenamefont {Nagahiro}, \citenamefont {Roca}, \citenamefont {Hosaka},\ and\ \citenamefont {Oset}}]{Nagahiro:2008cv}%
  \BibitemOpen
  \bibfield  {author} {\bibinfo {author} {\bibfnamefont {H.}~\bibnamefont {Nagahiro}}, \bibinfo {author} {\bibfnamefont {L.}~\bibnamefont {Roca}}, \bibinfo {author} {\bibfnamefont {A.}~\bibnamefont {Hosaka}}, \ and\ \bibinfo {author} {\bibfnamefont {E.}~\bibnamefont {Oset}},\ }\href {\doibase 10.1103/PhysRevD.79.014015} {\bibfield  {journal} {\bibinfo  {journal} {Phys. Rev. D}\ }\textbf {\bibinfo {volume} {79}},\ \bibinfo {pages} {014015} (\bibinfo {year} {2009})},\ \Eprint {http://arxiv.org/abs/0809.0943} {arXiv:0809.0943 [hep-ph]} \BibitemShut {NoStop}%
\bibitem [{\citenamefont {Zhang}\ and\ \citenamefont {Xie}(2018)}]{Zhang:2017eui}%
  \BibitemOpen
  \bibfield  {author} {\bibinfo {author} {\bibfnamefont {X.}~\bibnamefont {Zhang}}\ and\ \bibinfo {author} {\bibfnamefont {J.-J.}\ \bibnamefont {Xie}},\ }\href {\doibase 10.1088/0253-6102/70/1/60} {\bibfield  {journal} {\bibinfo  {journal} {Commun. Theor. Phys.}\ }\textbf {\bibinfo {volume} {70}},\ \bibinfo {pages} {060} (\bibinfo {year} {2018})},\ \Eprint {http://arxiv.org/abs/1712.05572} {arXiv:1712.05572 [nucl-th]} \BibitemShut {NoStop}%
\bibitem [{\citenamefont {Albrecht}\ \emph {et~al.}(1993)\citenamefont {Albrecht} \emph {et~al.}}]{ARGUS:1992olh}%
  \BibitemOpen
  \bibfield  {author} {\bibinfo {author} {\bibfnamefont {H.}~\bibnamefont {Albrecht}} \emph {et~al.} (\bibinfo {collaboration} {ARGUS}),\ }\href {\doibase 10.1007/BF01554080} {\bibfield  {journal} {\bibinfo  {journal} {Z. Phys. C}\ }\textbf {\bibinfo {volume} {58}},\ \bibinfo {pages} {61} (\bibinfo {year} {1993})}\BibitemShut {NoStop}%
\bibitem [{\citenamefont {Xie}\ \emph {et~al.}(2020)\citenamefont {Xie}, \citenamefont {Li},\ and\ \citenamefont {Liu}}]{Xie:2019iwz}%
  \BibitemOpen
  \bibfield  {author} {\bibinfo {author} {\bibfnamefont {J.-J.}\ \bibnamefont {Xie}}, \bibinfo {author} {\bibfnamefont {G.}~\bibnamefont {Li}}, \ and\ \bibinfo {author} {\bibfnamefont {X.-H.}\ \bibnamefont {Liu}},\ }\href {\doibase 10.1088/1674-1137/abae51} {\bibfield  {journal} {\bibinfo  {journal} {Chin. Phys. C}\ }\textbf {\bibinfo {volume} {44}},\ \bibinfo {pages} {114104} (\bibinfo {year} {2020})},\ \Eprint {http://arxiv.org/abs/1907.12202} {arXiv:1907.12202 [hep-ph]} \BibitemShut {NoStop}%
\bibitem [{\citenamefont {Xie}\ \emph {et~al.}(2023)\citenamefont {Xie}, \citenamefont {Lu}, \citenamefont {Geng},\ and\ \citenamefont {Zou}}]{Xie:2023cej}%
  \BibitemOpen
  \bibfield  {author} {\bibinfo {author} {\bibfnamefont {J.-M.}\ \bibnamefont {Xie}}, \bibinfo {author} {\bibfnamefont {J.-X.}\ \bibnamefont {Lu}}, \bibinfo {author} {\bibfnamefont {L.-S.}\ \bibnamefont {Geng}}, \ and\ \bibinfo {author} {\bibfnamefont {B.-S.}\ \bibnamefont {Zou}},\ }\href {\doibase 10.1103/PhysRevD.108.L111502} {\bibfield  {journal} {\bibinfo  {journal} {Phys. Rev. D}\ }\textbf {\bibinfo {volume} {108}},\ \bibinfo {pages} {L111502} (\bibinfo {year} {2023})},\ \Eprint {http://arxiv.org/abs/2307.11631} {arXiv:2307.11631 [hep-ph]} \BibitemShut {NoStop}%
\bibitem [{\citenamefont {Xie}\ \emph {et~al.}(2025)\citenamefont {Xie}, \citenamefont {Liu}, \citenamefont {Lu}, \citenamefont {Liang}, \citenamefont {Molina},\ and\ \citenamefont {Geng}}]{Xie:2025xew}%
  \BibitemOpen
  \bibfield  {author} {\bibinfo {author} {\bibfnamefont {J.-M.}\ \bibnamefont {Xie}}, \bibinfo {author} {\bibfnamefont {Z.-W.}\ \bibnamefont {Liu}}, \bibinfo {author} {\bibfnamefont {J.-X.}\ \bibnamefont {Lu}}, \bibinfo {author} {\bibfnamefont {H.}~\bibnamefont {Liang}}, \bibinfo {author} {\bibfnamefont {R.}~\bibnamefont {Molina}}, \ and\ \bibinfo {author} {\bibfnamefont {L.-S.}\ \bibnamefont {Geng}},\ }\href@noop {} {\  (\bibinfo {year} {2025})},\ \Eprint {http://arxiv.org/abs/2511.14380} {arXiv:2511.14380 [hep-ph]} \BibitemShut {NoStop}%
\bibitem [{\citenamefont {Roca}\ \emph {et~al.}(2007)\citenamefont {Roca}, \citenamefont {Hosaka},\ and\ \citenamefont {Oset}}]{Roca:2006am}%
  \BibitemOpen
  \bibfield  {author} {\bibinfo {author} {\bibfnamefont {L.}~\bibnamefont {Roca}}, \bibinfo {author} {\bibfnamefont {A.}~\bibnamefont {Hosaka}}, \ and\ \bibinfo {author} {\bibfnamefont {E.}~\bibnamefont {Oset}},\ }\href {\doibase 10.1016/j.physletb.2007.10.035} {\bibfield  {journal} {\bibinfo  {journal} {Phys. Lett. B}\ }\textbf {\bibinfo {volume} {658}},\ \bibinfo {pages} {17} (\bibinfo {year} {2007})},\ \Eprint {http://arxiv.org/abs/hep-ph/0611075} {arXiv:hep-ph/0611075} \BibitemShut {NoStop}%
\bibitem [{\citenamefont {Aceti}\ \emph {et~al.}(2015{\natexlab{a}})\citenamefont {Aceti}, \citenamefont {Dias},\ and\ \citenamefont {Oset}}]{Aceti:2015zva}%
  \BibitemOpen
  \bibfield  {author} {\bibinfo {author} {\bibfnamefont {F.}~\bibnamefont {Aceti}}, \bibinfo {author} {\bibfnamefont {J.~M.}\ \bibnamefont {Dias}}, \ and\ \bibinfo {author} {\bibfnamefont {E.}~\bibnamefont {Oset}},\ }\href {\doibase 10.1140/epja/i2015-15048-5} {\bibfield  {journal} {\bibinfo  {journal} {Eur. Phys. J. A}\ }\textbf {\bibinfo {volume} {51}},\ \bibinfo {pages} {48} (\bibinfo {year} {2015}{\natexlab{a}})},\ \Eprint {http://arxiv.org/abs/1501.06505} {arXiv:1501.06505 [hep-ph]} \BibitemShut {NoStop}%
\bibitem [{\citenamefont {Aceti}\ \emph {et~al.}(2015{\natexlab{b}})\citenamefont {Aceti}, \citenamefont {Xie},\ and\ \citenamefont {Oset}}]{Aceti:2015pma}%
  \BibitemOpen
  \bibfield  {author} {\bibinfo {author} {\bibfnamefont {F.}~\bibnamefont {Aceti}}, \bibinfo {author} {\bibfnamefont {J.-J.}\ \bibnamefont {Xie}}, \ and\ \bibinfo {author} {\bibfnamefont {E.}~\bibnamefont {Oset}},\ }\href {\doibase 10.1016/j.physletb.2015.09.068} {\bibfield  {journal} {\bibinfo  {journal} {Phys. Lett. B}\ }\textbf {\bibinfo {volume} {750}},\ \bibinfo {pages} {609} (\bibinfo {year} {2015}{\natexlab{b}})},\ \Eprint {http://arxiv.org/abs/1505.06134} {arXiv:1505.06134 [hep-ph]} \BibitemShut {NoStop}%
\bibitem [{\citenamefont {Xie}(2015)}]{Xie:2015wja}%
  \BibitemOpen
  \bibfield  {author} {\bibinfo {author} {\bibfnamefont {J.-J.}\ \bibnamefont {Xie}},\ }\href {\doibase 10.1103/PhysRevC.92.065203} {\bibfield  {journal} {\bibinfo  {journal} {Phys. Rev. C}\ }\textbf {\bibinfo {volume} {92}},\ \bibinfo {pages} {065203} (\bibinfo {year} {2015})},\ \Eprint {http://arxiv.org/abs/1509.06196} {arXiv:1509.06196 [nucl-th]} \BibitemShut {NoStop}%
\bibitem [{\citenamefont {Molina}\ \emph {et~al.}(2021)\citenamefont {Molina}, \citenamefont {Doering}, \citenamefont {Liang},\ and\ \citenamefont {Oset}}]{Molina:2021awn}%
  \BibitemOpen
  \bibfield  {author} {\bibinfo {author} {\bibfnamefont {R.}~\bibnamefont {Molina}}, \bibinfo {author} {\bibfnamefont {M.}~\bibnamefont {Doering}}, \bibinfo {author} {\bibfnamefont {W.~H.}\ \bibnamefont {Liang}}, \ and\ \bibinfo {author} {\bibfnamefont {E.}~\bibnamefont {Oset}},\ }\href {\doibase 10.1140/epjc/s10052-021-09574-y} {\bibfield  {journal} {\bibinfo  {journal} {Eur. Phys. J. C}\ }\textbf {\bibinfo {volume} {81}},\ \bibinfo {pages} {782} (\bibinfo {year} {2021})},\ \Eprint {http://arxiv.org/abs/2107.07439} {arXiv:2107.07439 [hep-ph]} \BibitemShut {NoStop}%
\bibitem [{\citenamefont {Shen}\ \emph {et~al.}(2025)\citenamefont {Shen}, \citenamefont {Geng},\ and\ \citenamefont {Xie}}]{Shen:2024jfr}%
  \BibitemOpen
  \bibfield  {author} {\bibinfo {author} {\bibfnamefont {Q.-H.}\ \bibnamefont {Shen}}, \bibinfo {author} {\bibfnamefont {L.-S.}\ \bibnamefont {Geng}}, \ and\ \bibinfo {author} {\bibfnamefont {J.-J.}\ \bibnamefont {Xie}},\ }\href {\doibase 10.1140/epja/s10050-025-01516-6} {\bibfield  {journal} {\bibinfo  {journal} {Eur. Phys. J. A}\ }\textbf {\bibinfo {volume} {61}},\ \bibinfo {pages} {42} (\bibinfo {year} {2025})},\ \Eprint {http://arxiv.org/abs/2409.05302} {arXiv:2409.05302 [hep-ph]} \BibitemShut {NoStop}%
\bibitem [{\citenamefont {G{\"u}lmez}\ \emph {et~al.}(2017)\citenamefont {G{\"u}lmez}, \citenamefont {Mei{\ss}ner},\ and\ \citenamefont {Oller}}]{Gulmez:2016scm}%
  \BibitemOpen
  \bibfield  {author} {\bibinfo {author} {\bibfnamefont {D.}~\bibnamefont {G{\"u}lmez}}, \bibinfo {author} {\bibfnamefont {U.~G.}\ \bibnamefont {Mei{\ss}ner}}, \ and\ \bibinfo {author} {\bibfnamefont {J.~A.}\ \bibnamefont {Oller}},\ }\href {\doibase 10.1140/epjc/s10052-017-5018-z} {\bibfield  {journal} {\bibinfo  {journal} {Eur. Phys. J. C}\ }\textbf {\bibinfo {volume} {77}},\ \bibinfo {pages} {460} (\bibinfo {year} {2017})},\ \Eprint {http://arxiv.org/abs/1611.00168} {arXiv:1611.00168 [hep-ph]} \BibitemShut {NoStop}%
\bibitem [{\citenamefont {Wang}\ and\ \citenamefont {Zou}(2021)}]{Wang:2021jub}%
  \BibitemOpen
  \bibfield  {author} {\bibinfo {author} {\bibfnamefont {Z.-L.}\ \bibnamefont {Wang}}\ and\ \bibinfo {author} {\bibfnamefont {B.-S.}\ \bibnamefont {Zou}},\ }\href {\doibase 10.1103/PhysRevD.104.114001} {\bibfield  {journal} {\bibinfo  {journal} {Phys. Rev. D}\ }\textbf {\bibinfo {volume} {104}},\ \bibinfo {pages} {114001} (\bibinfo {year} {2021})},\ \Eprint {http://arxiv.org/abs/2107.14470} {arXiv:2107.14470 [hep-ph]} \BibitemShut {NoStop}%
\bibitem [{\citenamefont {Wang}\ and\ \citenamefont {Zou}(2022)}]{Wang:2022pin}%
  \BibitemOpen
  \bibfield  {author} {\bibinfo {author} {\bibfnamefont {Z.-L.}\ \bibnamefont {Wang}}\ and\ \bibinfo {author} {\bibfnamefont {B.-S.}\ \bibnamefont {Zou}},\ }\href {\doibase 10.1140/epjc/s10052-022-10460-4} {\bibfield  {journal} {\bibinfo  {journal} {Eur. Phys. J. C}\ }\textbf {\bibinfo {volume} {82}},\ \bibinfo {pages} {509} (\bibinfo {year} {2022})},\ \Eprint {http://arxiv.org/abs/2203.02899} {arXiv:2203.02899 [hep-ph]} \BibitemShut {NoStop}%
\bibitem [{\citenamefont {Titov}\ \emph {et~al.}(2000)\citenamefont {Titov}, \citenamefont {Kampfer},\ and\ \citenamefont {Reznik}}]{Titov:2000bn}%
  \BibitemOpen
  \bibfield  {author} {\bibinfo {author} {\bibfnamefont {A.~I.}\ \bibnamefont {Titov}}, \bibinfo {author} {\bibfnamefont {B.}~\bibnamefont {Kampfer}}, \ and\ \bibinfo {author} {\bibfnamefont {B.~L.}\ \bibnamefont {Reznik}},\ }\href {\doibase 10.1007/s100500050427} {\bibfield  {journal} {\bibinfo  {journal} {Eur. Phys. J. A}\ }\textbf {\bibinfo {volume} {7}},\ \bibinfo {pages} {543} (\bibinfo {year} {2000})},\ \Eprint {http://arxiv.org/abs/nucl-th/0001027} {arXiv:nucl-th/0001027} \BibitemShut {NoStop}%
\bibitem [{\citenamefont {Titov}\ \emph {et~al.}(2002)\citenamefont {Titov}, \citenamefont {Kampfer},\ and\ \citenamefont {Reznik}}]{Titov:2001yw}%
  \BibitemOpen
  \bibfield  {author} {\bibinfo {author} {\bibfnamefont {A.~I.}\ \bibnamefont {Titov}}, \bibinfo {author} {\bibfnamefont {B.}~\bibnamefont {Kampfer}}, \ and\ \bibinfo {author} {\bibfnamefont {B.~L.}\ \bibnamefont {Reznik}},\ }\href {\doibase 10.1103/PhysRevC.65.065202} {\bibfield  {journal} {\bibinfo  {journal} {Phys. Rev. C}\ }\textbf {\bibinfo {volume} {65}},\ \bibinfo {pages} {065202} (\bibinfo {year} {2002})},\ \Eprint {http://arxiv.org/abs/nucl-th/0102032} {arXiv:nucl-th/0102032} \BibitemShut {NoStop}%
\bibitem [{\citenamefont {Dai}\ \emph {et~al.}(2020)\citenamefont {Dai}, \citenamefont {Roca},\ and\ \citenamefont {Oset}}]{Dai:2020vfc}%
  \BibitemOpen
  \bibfield  {author} {\bibinfo {author} {\bibfnamefont {L.~R.}\ \bibnamefont {Dai}}, \bibinfo {author} {\bibfnamefont {L.}~\bibnamefont {Roca}}, \ and\ \bibinfo {author} {\bibfnamefont {E.}~\bibnamefont {Oset}},\ }\href {\doibase 10.1140/epjc/s10052-020-8220-3} {\bibfield  {journal} {\bibinfo  {journal} {Eur. Phys. J. C}\ }\textbf {\bibinfo {volume} {80}},\ \bibinfo {pages} {673} (\bibinfo {year} {2020})},\ \Eprint {http://arxiv.org/abs/2005.02653} {arXiv:2005.02653 [hep-ph]} \BibitemShut {NoStop}%
\bibitem [{\citenamefont {L{\"u}}\ and\ \citenamefont {He}(2016)}]{Lu:2016nlp}%
  \BibitemOpen
  \bibfield  {author} {\bibinfo {author} {\bibfnamefont {P.-L.}\ \bibnamefont {L{\"u}}}\ and\ \bibinfo {author} {\bibfnamefont {J.}~\bibnamefont {He}},\ }\href {\doibase 10.1140/epja/i2016-16359-7} {\bibfield  {journal} {\bibinfo  {journal} {Eur. Phys. J. A}\ }\textbf {\bibinfo {volume} {52}},\ \bibinfo {pages} {359} (\bibinfo {year} {2016})},\ \Eprint {http://arxiv.org/abs/1603.04168} {arXiv:1603.04168 [hep-ph]} \BibitemShut {NoStop}%
\bibitem [{\citenamefont {Dai}\ \emph {et~al.}(2019)\citenamefont {Dai}, \citenamefont {Roca},\ and\ \citenamefont {Oset}}]{Dai:2018zki}%
  \BibitemOpen
  \bibfield  {author} {\bibinfo {author} {\bibfnamefont {L.~R.}\ \bibnamefont {Dai}}, \bibinfo {author} {\bibfnamefont {L.}~\bibnamefont {Roca}}, \ and\ \bibinfo {author} {\bibfnamefont {E.}~\bibnamefont {Oset}},\ }\href {\doibase 10.1103/PhysRevD.99.096003} {\bibfield  {journal} {\bibinfo  {journal} {Phys. Rev. D}\ }\textbf {\bibinfo {volume} {99}},\ \bibinfo {pages} {096003} (\bibinfo {year} {2019})},\ \Eprint {http://arxiv.org/abs/1811.06875} {arXiv:1811.06875 [hep-ph]} \BibitemShut {NoStop}%
\bibitem [{\citenamefont {Zeng}\ \emph {et~al.}(2020)\citenamefont {Zeng}, \citenamefont {Lu}, \citenamefont {Wang}, \citenamefont {Xie},\ and\ \citenamefont {Geng}}]{Zeng:2020och}%
  \BibitemOpen
  \bibfield  {author} {\bibinfo {author} {\bibfnamefont {C.-H.}\ \bibnamefont {Zeng}}, \bibinfo {author} {\bibfnamefont {J.-X.}\ \bibnamefont {Lu}}, \bibinfo {author} {\bibfnamefont {E.}~\bibnamefont {Wang}}, \bibinfo {author} {\bibfnamefont {J.-J.}\ \bibnamefont {Xie}}, \ and\ \bibinfo {author} {\bibfnamefont {L.-S.}\ \bibnamefont {Geng}},\ }\href {\doibase 10.1103/PhysRevD.102.076009} {\bibfield  {journal} {\bibinfo  {journal} {Phys. Rev. D}\ }\textbf {\bibinfo {volume} {102}},\ \bibinfo {pages} {076009} (\bibinfo {year} {2020})},\ \Eprint {http://arxiv.org/abs/2006.15547} {arXiv:2006.15547 [hep-ph]} \BibitemShut {NoStop}%
\bibitem [{\citenamefont {Bayar}\ \emph {et~al.}(2016)\citenamefont {Bayar}, \citenamefont {Aceti}, \citenamefont {Guo},\ and\ \citenamefont {Oset}}]{Bayar:2016ftu}%
  \BibitemOpen
  \bibfield  {author} {\bibinfo {author} {\bibfnamefont {M.}~\bibnamefont {Bayar}}, \bibinfo {author} {\bibfnamefont {F.}~\bibnamefont {Aceti}}, \bibinfo {author} {\bibfnamefont {F.-K.}\ \bibnamefont {Guo}}, \ and\ \bibinfo {author} {\bibfnamefont {E.}~\bibnamefont {Oset}},\ }\href {\doibase 10.1103/PhysRevD.94.074039} {\bibfield  {journal} {\bibinfo  {journal} {Phys. Rev. D}\ }\textbf {\bibinfo {volume} {94}},\ \bibinfo {pages} {074039} (\bibinfo {year} {2016})},\ \Eprint {http://arxiv.org/abs/1609.04133} {arXiv:1609.04133 [hep-ph]} \BibitemShut {NoStop}%
\end{thebibliography}%
\end{document}